\documentclass[10pt,journal]{IEEEtran}
\usepackage{tabularx}
\usepackage[T1]{fontenc}
\usepackage{algorithmic}
\usepackage{graphicx}
\usepackage{amsmath,amssymb,amsfonts,stfloats}
\usepackage{array}
\usepackage{textcomp}
\usepackage[font=small]{caption}
\usepackage{paralist}
\usepackage{lettrine}
\usepackage{cite}
\usepackage{xcolor}
\usepackage{cancel}
\usepackage{mathtools}
\usepackage{booktabs}
\usepackage[caption=false,font=footnotesize]{subfig}
\usepackage{url}
\usepackage{acronym}
\usepackage{multicol}
\usepackage[ruled,vlined]{algorithm2e}

\begin{document}

\title{Wavefield Networked Sensing: \\Principles, Algorithms and Applications 
\thanks{The authors are with Politecnico di Milano, 20133 Milano, Italy, Department of Electronics, Information and Bioengineering. Emails: [name.surname]@polimi.it}\thanks{This work was partially supported by the European Union under the Italian National Recovery and Resilience Plan (NRRP) of NextGenerationEU, partnership on “Telecommunications of the Future” (PE00000001 - program “RESTART”) }}

\author{Marco Manzoni, Dario Tagliaferri, Stefano Tebaldini, Marouan Mizmizi, \\ Andrea~Virgilio Monti-Guarnieri, Claudio Maria Prati and Umberto Spagnolini}
\maketitle

\begin{abstract}
Networked sensing refers to the capability of properly orchestrating multiple sensing terminals to enhance specific figures of merit, e.g., positioning accuracy or imaging resolution. Regarding radio-based sensing, it is essential to understand \textit{when} and \textit{how} sensing terminals should be orchestrated, namely the best cooperation that trades between performance and cost (e.g., energy consumption, communication overhead, and complexity). This paper addresses networked sensing from a physics-driven perspective, aiming to provide a general theoretical benchmark to evaluate its \textit{imaging} performance bounds and to guide the sensing orchestration accordingly. Diffraction tomography theory (DTT) is the method to quantify the imaging resolution of any radio sensing experiment from inspection of its spectral (or wavenumber) content. In networked sensing, the image formation is based on the back-projection integral, valid for any network topology and physical configuration of the terminals. The \textit{wavefield networked sensing} is a framework in which multiple sensing terminals are orchestrated during the acquisition process to maximize the imaging quality (resolution and grating lobes suppression) by pursuing the deceptively simple \textit{wavenumber tessellation principle}. We discuss all the cooperation possibilities between sensing terminals and possible killer applications. Remarkably, we show that the proposed method allows obtaining high-quality images of the environment in limited bandwidth conditions, leveraging the coherent combination of multiple multi-static low-resolution images.

\end{abstract}

\begin{IEEEkeywords}
Networked sensing, diffraction tomography theory, imaging, orchestration 
\end{IEEEkeywords}

\section{Introduction}\label{sect:introduction}
Sensing is a fundamental task performed in many applications as the catalyst for context awareness, such as autonomous driving, industrial automation, smart cities, and smart homes. It enables devices to gather data from the environment and provide valuable insights for decisions \cite{temdee2018context}. Sensing can be performed using different technologies depending on the specific application, i.e., radio, light, and vision, each with benefits and challenges \cite{rosique2019systematic, 7870764, 9455394}. For instance, radio-based sensing can be used for long-range applications and can penetrate walls, while light-based sensing is suitable for short-range applications and requires line-of-sight. In radio-based sensing, which is the focus of this paper, a notable approach to improving sensing performance and reliability w.r.t. the single terminal is to use multiple terminals, following the paradigm of \textit{networked sensing}. However, as energy consumption and complexity rise quickly with the number of sensing terminals, the effort may not be worthwhile. Consequently, it is essential to predict networked sensing performance and assess the cost of achieving it. In the following we discuss the networked sensing fundamental concepts and provide a list of possible applications, with the paper contribution.
%

\subsection{Networked Sensing Fundamentals}

Consider the case in which multiple sensing terminals, possibly with different capabilities, are willing to cooperate to enhance the sensing quality with respect to the single terminal. Terminals can be any, e.g., infrastructure, users, intelligent vehicles, etc. The fundamental questions are: 
\begin{itemize}
    \item[\textbf{Q1}] \textbf{What is the desired output of sensing?}
    \item[\textbf{Q2}] \textbf{What is the most appropriate key performance indicator for networked sensing?}
    \item[\textbf{Q3}] \textbf{What are the practical options for cooperation in networked sensing?}
\end{itemize}
Questions Q1 and Q2 are related to the choice of the desired "product" of sensing in a given application scenario, with related figures of merit. On one side, sensing can be used to \textit{localize} the targets in the environment, namely estimating their position, orientation, and possibly velocity, irrespective of their physical shape/size. For instance, localizing a vehicle can be intended as the estimation of the position/velocity of the communication transceiver, as far as communication is concerned. A common valid performance metric for localization, almost always adopted, is the Cramér-Rao lower bound (CRLB) on position/velocity/orientation estimation. CRLB is tightly conditioned to a given signal and noise model, and it is generally cumbersome to be obtained in closed form for more than one target, as Fisher's information matrix size scales quadratically with the number of targets \cite{Kay}. Therefore, CRLB is inherently suited to relatively sparse scenarios, with few \textit{weakly coupled} (or not coupled) targets, i.e., the estimation of the parameters of each is not affected by the presence of the others \cite{Chetty2022_CRB}. On the other hand, when the interest is in the whole environment mapping, and there are multiple classes of targets, \textit{imaging} is far more appropriate. Imaging a target means adding the information on its shape and size, i.e., creating a 2D or 3D map of the complex reflectivity of the environment, from which the estimation of position/velocity/shape of targets follows after detection. In this framework, the assumption of weakly coupled targets is no more valid and the CRLB is not the best option to evaluate the sensing performance, which is instead measured using other metrics that are typically used to qualify radar systems imaging performances such as resolution, peak-to-sidelobe ratio (PSLR), integrated sidelobe ratio (ISLR) \cite{cumming}. 

Question Q3 aims at understanding \textit{when} and \textit{how} the cooperation between terminals can take place, with a given resource budget, i.e., time-frequency-space-hardware and sensing network topology. Cooperation can be either \textit{loose}, (i.e., terminals perform independent measurements to be mutually exchanged and properly fused to improve one specific figure of merit sich as detection probability) or \textit{tight}, (i.e., terminals act like a distributed sensor). Moreover, sensing data from different terminals can be fused \textit{coherently} or \textit{incoherently}, thus with or without retaining the phase information in the measured data during the fusion, whereas measurements can be acquired in a \textit{monostatic} setting (Tx and Rx coincide) and/or in a \textit{multistatic} setting (Tx and Rx do not coincide). The coherent tight fusion of sensing data allows exploring the utter sensing performance, where terminals perform a joint acquisition, properly scheduled to maximize a given figure of merit of the sensing output. In general, the extent of the cooperation is function of \textit{(i)} the number of available sensing terminals and related spatial density, \textit{(ii)} the terminals' sensing capabilities and \textit{(iv)} the trade-off between performance improvement (resolution, signal-to-noise ratio (SNR), etc.) and cost (e.g., spare communication resources for the cooperation, effort in terminals synchronization).

\subsection{Application Scenarios and Literature Survey}
Networked sensing is the enabling technology in several application scenarios supporting existing and novel verticals, from remote sensing of the environment to 6th generation of communication systems (6G). Some of them are listed in the following with the related literature.

\subsubsection{Networked sensing in 6G integrated sensing and communication (ISAC) systems} 
ISAC systems are surging as one of the key technologies for 6G systems, evolving from "listen and talk" paradigm to "see and feel" \cite{Liu_survey}. Given the envisioned density of ISAC terminals in space, (i.e, base stations (BS) and users), networked sensing promises to enhance ISAC performance by fusing multiple views of the same environment while handling the mutual interference among terminals. Existing literature on networked ISAC is limited to very recent works. The paper \cite{Eldar2022} reports an overview of the potential and challenges of networked ISAC, spotting the issues of interference management. The work \cite{Huang2022_powercontrol} proposes a proper power control to maximize the SNR while minimizing the CRLB on position estimation. The work in \cite{Shi2022_DEVICEFREE} considers the problem of fusing multiple monostatic sensing measurements of the same target from different ISAC nodes in order to reduce the issue of ghost targets. Information theory is used in \cite{Caire2022_InfTh_JCS} to stress the potential of networked ISAC systems with multiple receiving terminals. As the natural extension of cellular systems, cell-free MIMO systems aim at substituting massive MIMO BSs with many low-cost \textit{access points} (AP), equipped with few antennas \cite{9336188}. A single communication user is typically served by multiple APs, that shall emit the signal to guarantee that it adds up in phase at the user location. In this setting, the benefit of networked sensing is evident \cite{demirhan2023cellfree}.

\subsubsection{High-fidelity digital twins (DTs) for Autonomous Driving} A DT is a high-fidelity digital representation of the physical environment, obtained by merging simulations and available measured data in real-time to keep the DT updated \cite{Mihai_DT_survey,Ding_DT_comm}. DT exploits multiple measurements from heterogeneous sensors (radar, communication, camera, lidars, and others) from multiple sensing terminals, e.g., ISAC nodes, pedestrians, vehicles, etc. In particular, autonomous vehicles perceive the environment and exchange the measurements to create, update and augment the DT, which is then made available to everyone to realize the collective perception of the environment \cite{Schwarz_DT_AD}. In the autonomous driving context, networked sensing in the radio spectrum can increase the robustness of autonomous platoons, complementing optical sensors in adverse weather conditions with high-accuracy maps of the environment \cite{Marti2019ADAS_sensors}. 

\subsubsection{Non-Terrestrial Networks (NTN)}
Remote sensing was one of the first use-cases for non-terrestrial terminals, (i.e., satellites), exploiting mainly radio and vision technologies for different applications, e.g., weather, surveillance, and cartography \cite{6327381,7974728}. The usage of NTN to have spatial diversity is known for a long time. Focusing only on satellite missions for brevity, tomography with synthetic aperture radar (SAR) systems uses multiple satellites passes to map the Earth with fine resolution, allowing the retrieval of forest parameters such as height or biomass \cite{HoTongMinh2014, Blomberg2021, Tebaldini2023, MariottiDAlessandro2019} as well as the mapping of glacier and their internal structure \cite{Tebaldini2012, Frey2016, Tebaldini2016, Banda2016}. Interferometric techniques, leveraging the radar measurements from two orbits, are also used to generate digital elevation models (DEM) of the Earth \cite{Zink2021}. Recent technological advancements unlocked many novel applications of NTNs for 6G systems, especially when integrated with terrestrial networks \cite{9861699}. In this context, networked sensing refers to the capability of allocating resources (acquisition time, frequency, and orbits) to maximize a given figure of merit.

\subsection{Paper Contribution}

This paper addresses the aforementioned networked sensing challenges (Q1,Q2 and Q3) from the imaging perspective. In the remote sensing context, imaging capabilities are often formalized by invoking diffraction tomography theory (DTT) \cite{WuToksoz1987}. This concept is derived from the electromagnetic wave theory of scattering and it essentially states that the response of a target to a collection of monochromatic waves is the Fourier transform of its spatial reflectivity function. 
Any 2D radio sensing image can be expressed either in conventional spatial coordinates or in the dual wavenumber domain, where wavenumbers are here intended as the spatial angular frequencies. Spatial resolution is among those properties that appear particularly well in the wavenumber domain. Indeed, it is a general principle that the finer the spatial resolution of a given image, the larger the region in the wavenumber domain covered by the Fourier transform (FT) of the same image. As an example, this principle is used in \cite{Tebaldini2017_tessellation} to design the orbits of a swarm of spaceborne radars with the aim of achieving fine spatial resolution while transmitting narrow-bandwidth signals. Our goal is to generalize \cite{Tebaldini2017_tessellation} to \textit{any} possible scenario and use case, demonstrating the potential of DTT as a fundamental theory to evaluate the networked sensing performance in terms of imaging for arbitrary geometries and acquisition parameters (terminals' capabilities). We convey the concept of \textit{wavefield coherent networked sensing}, where multiple sensing terminals are properly orchestrated to maximize the imaging quality following the \textit{wavenumber tessellation principle}, and the result of the orchestration is interpreted from a physics-driven perspective.

The main contributions are the following:

\begin{itemize}
    \item We introduce and discuss the DTT, deriving the closed-form expression of the wavenumber covered region in many practical use cases with arbitrary geometry (monostatic vs. bistatic/multi-static);
    \item We outline and discuss the back-projection integral, which is the most general image formation algorithm and can be used in any geometrical setting and physical capabilities of the sensors (near-field and far-field, narrowband and wideband operation). The back-projection algorithm is conceptually derived starting from the DTT, linking the wavenumber-covered region of a given networked sensing experiment with the corresponding imaging capabilities;
    \item We discuss the advantages of different levels of cooperation among the sensing terminals, from the basic exchange of amplitude-only and amplitude-plus-phase data acquired at each terminal (incoherent vs. coherent sum of monostatic images) to the full orchestrated sensing in which each terminal is scheduled to perform both monostatic and multistatic acquisition to explicitly enhance the imaging resolution. In particular, the latter is a further step w.r.t. state of the art, as it allows to design of the guidelines for networked sensing by pursuing the wavenumber tessellation principle. Requirements for clock synchronization are also outlined and briefly discussed.
    
    \item Finally, we show a design guideline for networked sensing where cooperation is not recommended, and we outline some future perspectives on promising application examples and research directions. 
\end{itemize}

\subsection{Organization and Notation}
The paper is organized as follows: Section \ref{sect:FEDT} describes the DTT, Section \ref{sec:focusing_algorithms} derives the back-projection integral from the DTT as image formation algorithm, Section \ref{sec:options_for_cooperation} discusses the possible options for cooperation between sensing terminals, while Section \ref{sec:design_guideline} reports an example of design guideline for networked imaging systems. Finally, Section \ref{sect:conclusion} concludes the paper and sheds light on possible future research perspectives.

The following notation is adopted: bold upper- and lower-case denote matrices and column vectors. The L2-norm of a vector is denoted with $\|\cdot\|$. Matrix transposition is indicated as $\mathbf{A}^T$. $\mathbf{I}_n$ is the identity matrix of size $n$.  $\mathbf{a}\sim\mathcal{CN}(\boldsymbol{\mu},\mathbf{C})$ denotes a multi-variate circularly complex Gaussian random variable with mean $\boldsymbol{\mu}$ and covariance $\mathbf{C}$. $\mathbb{R}$ and $\mathbb{C}$ stand for the set of real and complex numbers, respectively. $\delta_{n}$ is the Kronecker delta.

\section{The Wavenumber Domain and the DTT}
\label{sect:FEDT}
The theoretical background of the present work lies in the diffraction tomography theory \cite{WuToksoz1987}. Any 2D radio sensing image can be expressed either in conventional spatial coordinates $\mathbf{x}=[x,y]^\mathrm{T}$ or in the dual wavenumber domain $\mathbf{k}=[k_x,k_y]^\mathrm{T}$. Extensions to 3D scenarios are possible but not covered here. In the following, we briefly recall the DTT, which allows calculating the region in the wavenumber domain that is covered by a sensing acquisition given the transmitted carrier frequency, bandwidth and the positions of the Tx and the Rx sensors, relative to the scene.
\begin{figure}[!t]
    \centering
    \subfloat[]{\includegraphics[width=\columnwidth]{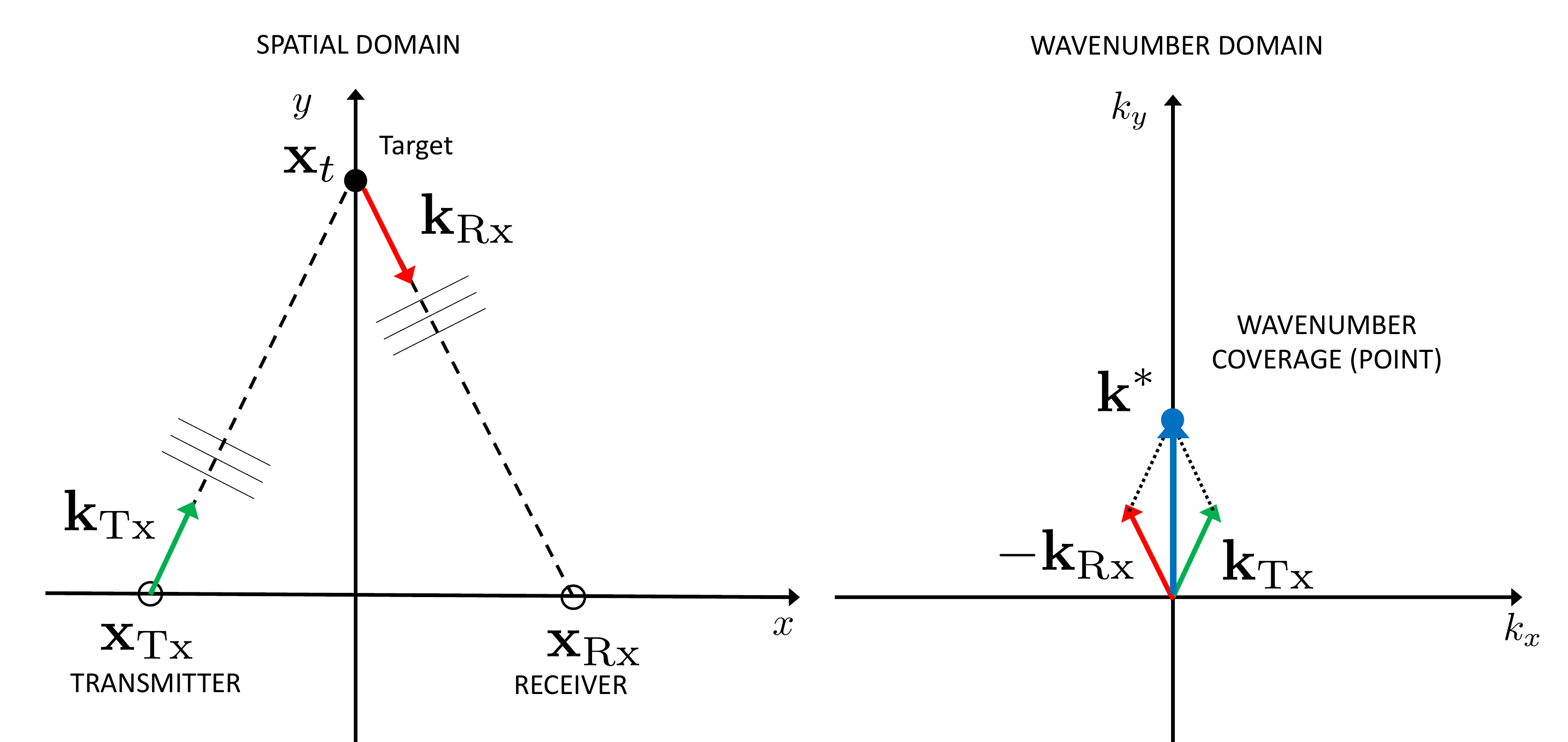}\label{subfig:w1}}\\
    \subfloat[]{\includegraphics[width=\columnwidth]{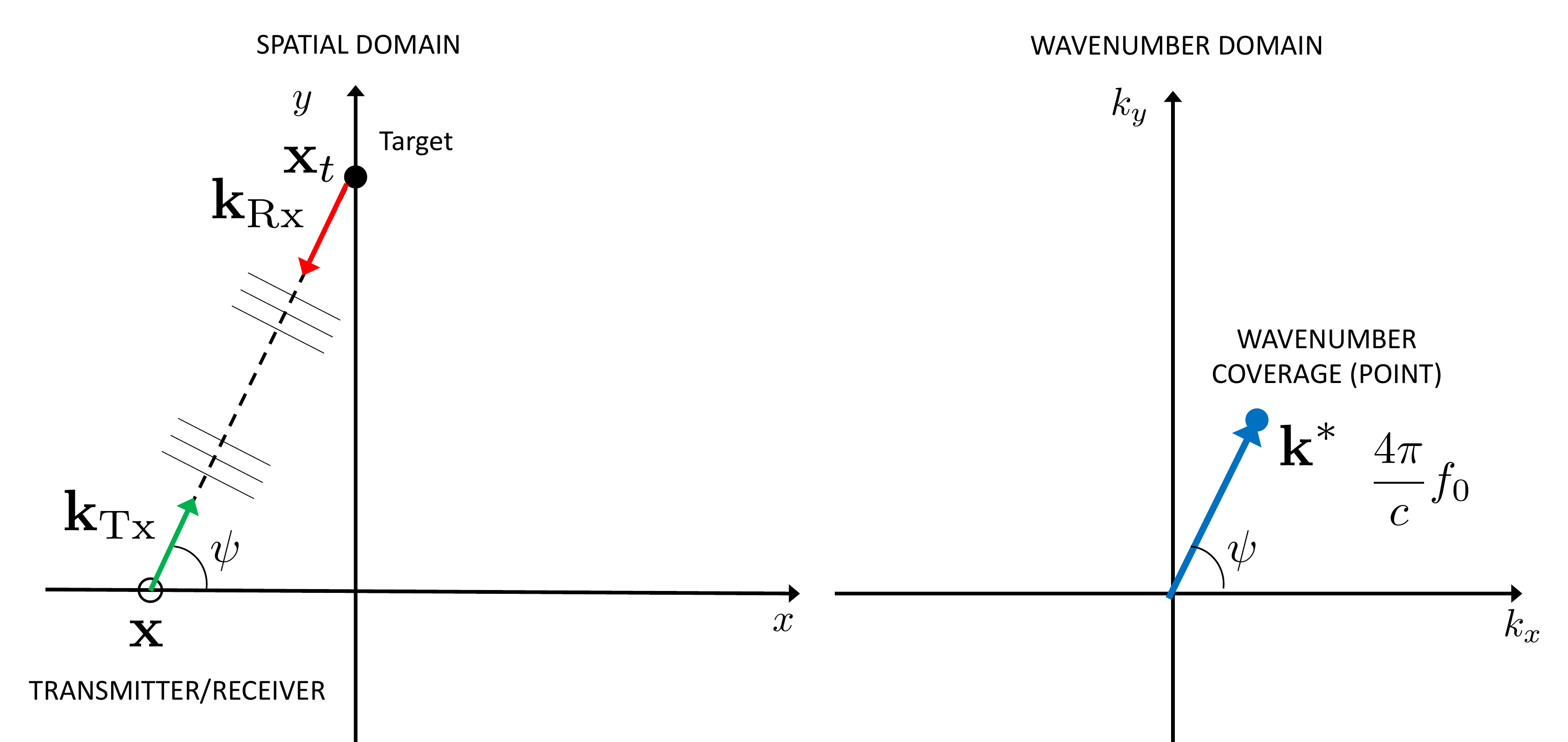}\label{subfig:w2}}
    \caption{Illustration of the wavenumber covered region for: generic monochromatic radar acquisition (\ref{subfig:w1}), monostatic mono-frequency radar acquisition (\ref{subfig:w2}).}
    \label{fig:wavenumbers}
\end{figure}

\begin{figure}[!t]
    \centering
    \subfloat[]{\includegraphics[width=\columnwidth]{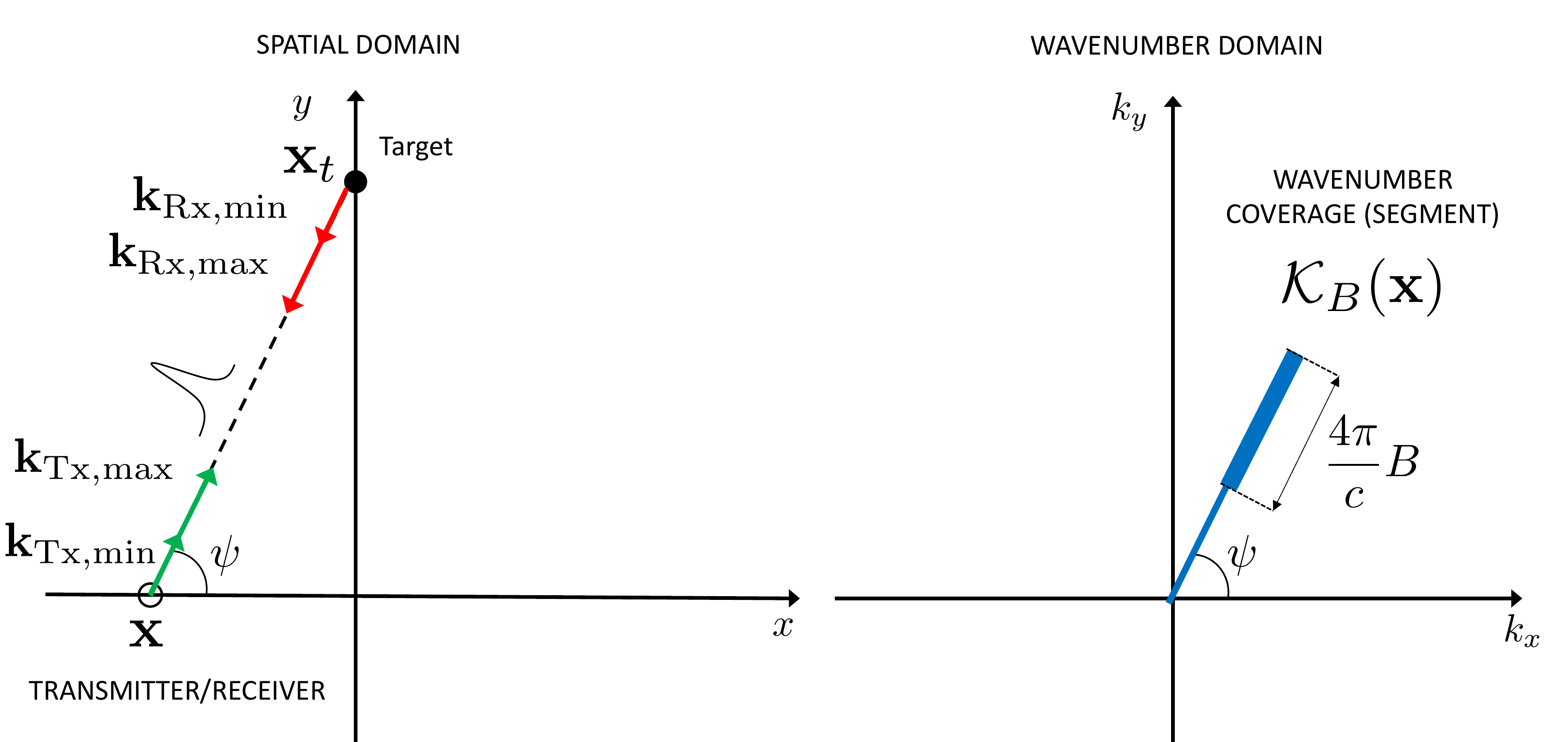}\label{subfig:w3}}\\
    \subfloat[]{\includegraphics[width=\columnwidth]{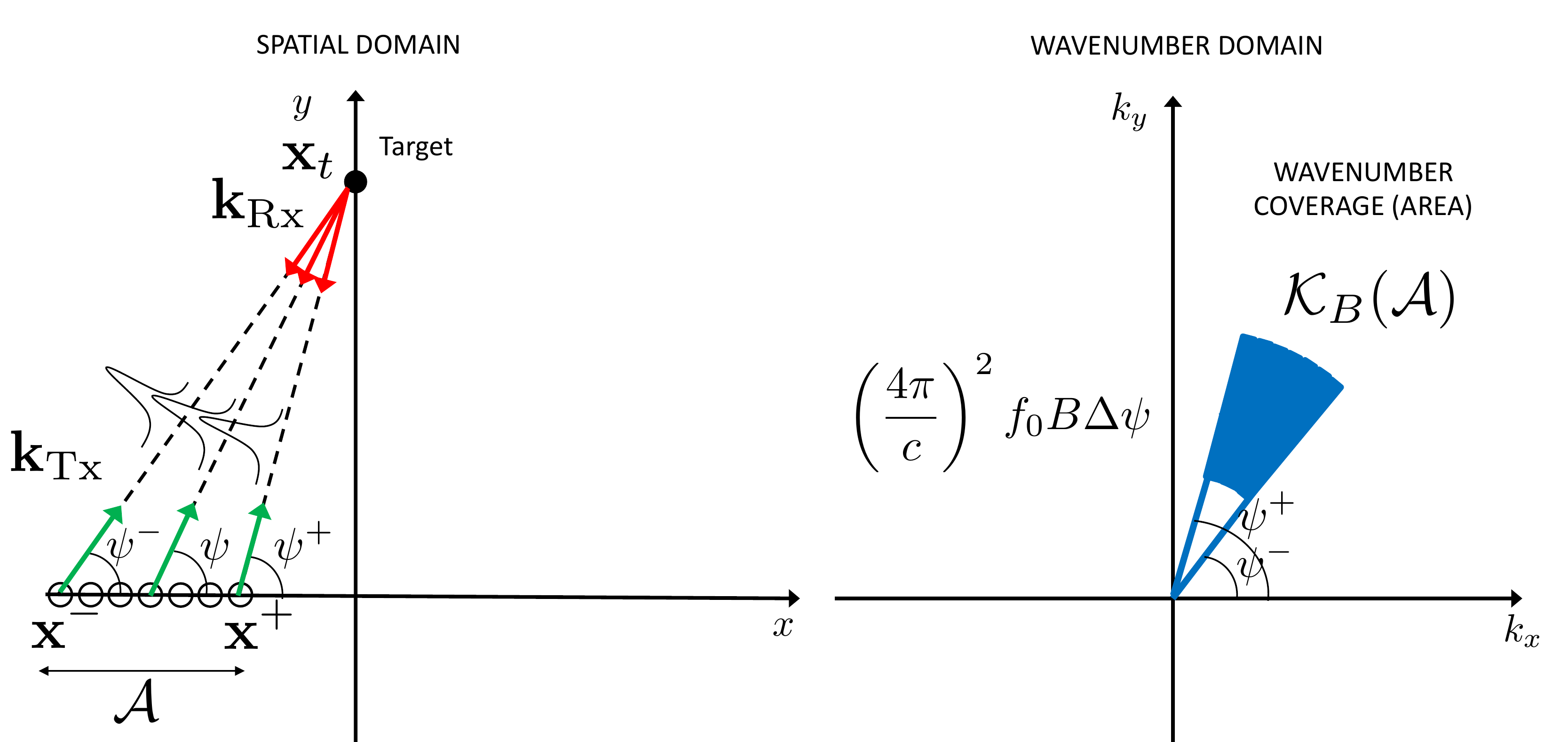}\label{subfig:w4}}\\
    \caption{Illustration of the wavenumber covered region for: monostatic bandwidth-limited radar acquisition (\ref{subfig:w3}) 
    monostatic bandwidth-limited radar acquisition with aperture (\ref{subfig:w4}) 
    }
    \label{fig:wavenumbers2}
\end{figure}
%
Consider a generic 2D radio sensing experiment, portrayed in Fig. \ref{subfig:w1}, where a perfectly isotropic target in $\mathbf{x}_t = [x_t,y_t]^\mathrm{T}$ is illuminated by a monochromatic plane wave at frequency $f_0$ emitted by a transmitter located at $\mathbf{x}_{\mathrm{Tx}} = [x_{\mathrm{Tx}},y_{\mathrm{Tx}}]^\mathrm{T}$, and the reflected wave is then recorded by a receiver at $\mathbf{x}_{\mathrm{Rx}}=[x_{\mathrm{Rx}},y_{\mathrm{Rx}}]^\mathrm{T}$. Neglecting geometrical energy losses and noise, which are irrelevant for this description, the received signal can be modeled as:
\begin{equation}\label{eq:Rxsignal_pointtarget}
    r(\mathbf{x}_{\mathrm{Tx}},\mathbf{x}_{\mathrm{Rx}}) = \gamma(\mathbf{x}_t) e^{-j k_0 (R_{\mathrm{Tx}}(\mathbf{x}_t) + R_{\mathrm{Rx}}(\mathbf{x}_t))}
\end{equation}
where $k_0 = 2\pi f_0/c$ is the wavenumber (with $c=3\times 10^8$ m/s being the speed of light in vacuum), $R_{\mathrm{Tx}}(\mathbf{x}_t) = \|\mathbf{x}_t - \mathbf{x}_\mathrm{Tx}\|$ and $R_{\mathrm{Rx}}(\mathbf{x}_t)=\|\mathbf{x}_{\mathrm{Rx}}-\mathbf{x}_t\|$ are the distances between the Tx and Rx and the target, respectively, $\gamma(\mathbf{x}_t)\in\mathbb{C}$ is the complex reflectivity of the target, modeling the interaction between the impinging wave and the target. The reflectivity $\gamma(\mathbf{x}_t)$ can also be interpreted as the reflectivity map of the scene $\gamma(\mathbf{x})$ evaluated in $\mathbf{x}=\mathbf{x}_t$.
If the weak scattering condition holds (see \cite{Wolf1969} for further details), we can extend \eqref{eq:Rxsignal_pointtarget} to an arbitrarily shaped target as:
\begin{equation}\label{eq:Rxsignal_exttarget}
    \begin{split}
r(\mathbf{x}_{\mathrm{Tx}},\mathbf{x}_{\mathrm{Rx}}) & \approx   \iint_S \gamma(\mathbf{x}) e^{-j k_0 (R_{\mathrm{Tx}}(\mathbf{x}) + R_{\mathrm{Rx}}(\mathbf{x}))} \mathrm{d}\mathbf{x}
    \end{split}
\end{equation}
where the integral extends over the visible area $S$ of the target.
Let us consider a conveniently small neighborhood region around $\mathbf{x}_t$, and write distances as a first-order Taylor expansion as:
\begin{equation}\label{eq:distances}
\begin{split}
     R_\mathrm{Tx}(\mathbf{x}) & \approx R_\mathrm{Tx}(\mathbf{x}_t) + \nabla R_\mathrm{Tx}(\mathbf{x}_t)^{\mathrm{T}} (\mathbf{x} - \mathbf{x}_t)\\
     R_\mathrm{Rx}(\mathbf{x}) & \approx R_\mathrm{Rx}(\mathbf{x}_t) + \nabla R_\mathrm{Rx}(\mathbf{x}_t)^{\mathrm{T}} (\mathbf{x} - \mathbf{x}_t)
\end{split}
\end{equation}
where $\left\|\nabla R(\mathbf{x}_t)\right\|=1$ by construction. Equation \eqref{eq:Rxsignal_exttarget} becomes
\begin{equation}\label{eq:Rxsignal_exttarget_simple}
\begin{split}
r(\mathbf{x}_{\mathrm{Tx}},\mathbf{x}_{\mathrm{Rx}}) & \approx e^{-j k_0 (R_{\mathrm{Tx}}(\mathbf{x}_t) + R_{\mathrm{Rx}}(\mathbf{x}_t))} \times \\ & \times \iint_S \gamma(\mathbf{x}) e^{-j k_0 (\nabla R_{\mathrm{Tx}}(\mathbf{x}_t) + \nabla R_{\mathrm{Rx}}(\mathbf{x}_t))^{\mathrm{T}} \mathbf{x}}\;\mathrm{d}\mathbf{x}.
\end{split}
\end{equation}
Vectors
\begin{equation}\label{eq:wavevectors}
    \begin{split}
        \mathbf{k}_{\mathrm{Tx}} & = k_0 \nabla R_{\mathrm{Tx}}(\mathbf{x}_t) = \frac{2\pi f_0}{c}[\cos\psi_{\mathrm{Tx}}, \sin\psi_{\mathrm{Tx}}]^\mathrm{T}\\
        \mathbf{k}_{\mathrm{Rx}} & = - k_0 \nabla R_{\mathrm{Rx}}(\mathbf{x}_t)= -\frac{2\pi f_0}{c}[\cos\psi_{\mathrm{Rx}}, \sin\psi_{\mathrm{Rx}}]^\mathrm{T}
    \end{split}
\end{equation}
denote plane wavevectors from Tx to the target and from the target to Rx, respectively, while $\psi_{\mathrm{Tx}}$ and $\psi_{\mathrm{Rx}}$ are the Tx and Rx observation angles, expressed in global coordinates. 

We can recognize in \eqref{eq:Rxsignal_exttarget_simple} the 2D Fourier transform of the target's reflectivity evaluated in 
\begin{equation}
\label{eq:enlight_wavenumber}
    \mathbf{k}^* = \mathbf{k}_{\mathrm{Tx}}-\mathbf{k}_{\mathrm{Rx}}
\end{equation}
 leading to:
\begin{equation}\label{eq:Rxsignal_FT}
    r(\mathbf{x}_{\mathrm{Tx}},\mathbf{x}_{\mathrm{Rx}}) \approx e^{-j k_0 (R_{\mathrm{Tx}}(\mathbf{x}_t) + R_{\mathrm{Rx}}(\mathbf{x}_t))} \Gamma(\mathbf{k}^*) 
\end{equation}
where 
\begin{equation}
    \Gamma(\mathbf{k}) = \iint_S \gamma(\mathbf{x})e^{-j \mathbf{k}^{\mathrm{T}}\mathbf{x}}\; \mathrm{d}\mathbf{x}
\end{equation}
denotes the FT of the target's reflectivity $\gamma(\mathbf{x})$. The theory developed in this Section provides a remarkable interpretation of radio sensing images. Any radio sensing measurement, indeed, simply \textit{excites} some target's wavenumbers $\mathbf{k}^*$ according to \eqref{eq:enlight_wavenumber}. When the relative position between the sensor and the target changes, different wavenumbers are excited leading to finer resolution, in some cases.
Notice that for a perfect isotropic target (a point) we have:
\begin{equation}
  \gamma(\mathbf{x})=\gamma_t \, \delta(\mathbf{x}-\mathbf{x}_t) \rightarrow  \lvert\Gamma(\mathbf{k})\rvert = 1
\end{equation}
i.e., the FT spans the infinite wavenumber domain. In this condition, the scene to be imaged is said to be \textit{white}. Choosing a point target is a common practice in the sensing context, as the wavenumber covered region of $\Gamma(\mathbf{k})$ (also known as \textit{spectral support}) depends solely on the specific sensing experiment and not on the shape of the target. In other words, a single point target represents the most challenging target and it allows analyzing the \textit{spatial impulse response function} of a given scene, for an arbitrary acquisition geometry. The response to more complex targets is simply the collection of responses to multiple point targets, for the linearity of the Fourier transform.


%
\subsection{Monostatic acquisition ($\mathbf{x}_\mathrm{Tx}=\mathbf{x}_\mathrm{Rx}=\mathbf{x}$)}\label{subsect:monostatic_wavenumber_region}

Let us consider a monostatic radar acquisition, where $\mathbf{x}_\mathrm{Tx}=\mathbf{x}_\mathrm{Rx} = \mathbf{x}$. If the wave is monochromatic at frequency $f_0$ (Fig. \ref{subfig:w2}), we can only observe a single wavenumber $\mathbf{k}^*$ of the scene, such that $\widehat{\Gamma}(\mathbf{k})=\delta(\mathbf{k}-\mathbf{k}^*)$, where $\|\mathbf{k}^*\| = 4 \pi f_0/c$. The direct consequence of the absence of observed bandwidth is the absence of resolution, thus the position of the target is unknown.
On the contrary, by transmitting an impulse of bandwidth $B$ centered around $f_0$, we add range resolution to the imaging system. Fig. \ref{subfig:w3} show the same monostatic experiment of Fig. \ref{subfig:w2}, where the Tx and Rx wavevectors range from $\mathbf{k}_{\mathrm{Tx,min}}$ to $\mathbf{k}_{\mathrm{Tx,max}}$ (Tx),  and from $\mathbf{k}_{\mathrm{Tx,min}}$ to $\mathbf{k}_{\mathrm{Tx,max}}$ (Rx), whose expressions are:
\begin{align}
    \mathbf{k}_{\mathrm{Tx,min}} & = \frac{2\pi(f_0 + f_\mathrm{min})}{c} [\cos\psi_{\mathrm{Tx}}, \sin\psi_{\mathrm{Tx}}]^\mathrm{T} \label{eq:wavenumber_min_Tx}\\
    \mathbf{k}_{\mathrm{Tx,max}} & = \frac{2\pi(f_0 + f_\mathrm{max})}{c} [\cos\psi_{\mathrm{Tx}}, \sin\psi_{\mathrm{Tx}}]^\mathrm{T} \label{eq:wavenumber_max_Tx}\\
    \mathbf{k}_{\mathrm{Rx,min}} & = -\frac{2\pi(f_0 + f_\mathrm{min})}{c} [\cos\psi_{\mathrm{Rx}}, \sin\psi_{\mathrm{Rx}}]^\mathrm{T} \label{eq:wavenumber_min_Rx}\\
    \mathbf{k}_{\mathrm{Rx,max}} & = -\frac{2\pi(f_0 + f_\mathrm{max})}{c} [\cos\psi_{\mathrm{Rx}}, \sin\psi_{\mathrm{Rx}}]^\mathrm{T} \label{eq:wavenumber_max_Rx}
\end{align}
where $f_\mathrm{min} = -B/2$ and $f_\mathrm{max} = B/2$ are the limit of the the base-band spectrum of the Tx pulse. The covered region in the wavenumber domain is now the segment
\begin{equation}\label{eq:wavenumber_region_segment}
    \mathcal{K}_{B}(\mathbf{x}) = \left\{ \mathbf{k} = \eta\,\mathbf{k}^*_{\mathrm{min}}+(1-\eta)\mathbf{k}^*_{\mathrm{max}},\,0\leq\eta\leq 1\right\}
\end{equation}
spanned by wavevectors $\mathbf{k}^*_{\mathrm{min}} = \mathbf{k}_{\mathrm{Tx,min}}-\mathbf{k}_{\mathrm{Rx,min}}$ and $\mathbf{k}^*_{\mathrm{max}} = \mathbf{k}_{\mathrm{Tx,max}}-\mathbf{k}_{\mathrm{Rx,max}}$. The segment \eqref{eq:wavenumber_region_segment} has length $\lvert\mathcal{K}_{B}(\mathbf{x})\rvert = 4\pi B/c$ along the radial direction and it is centered in $4\pi f_0/c$. 
%
%
%
%
%
The range resolution of the reconstructed image is:
\begin{equation}
\label{eq:range_resolution}
    \rho_r = \frac{2 \pi}{\lvert\mathcal{K}_{B}(\mathbf{x})\rvert} =  \frac{c}{2B}.
\end{equation} 
Let us consider now the combination of multiple monostatic acquisitions, performed along the \textit{aperture} in space
\begin{equation}\label{eq:aperture}
    \mathcal{A} = \left\{ \mathbf{x} = \mu \mathbf{x}^- + (1-\mu)\mathbf{x}^+, 0\leq\mu\leq 1 \right\}
\end{equation}
of length $\|\mathbf{x}^+ - \mathbf{x}^-\|=A$ (Fig. \ref{subfig:w4}), where $\mathbf{x}^-$ and $\mathbf{x}^+$ are the two segment ending points. The aperture in \eqref{eq:aperture} can represent a physical antenna array (at each real or virtual channel) or synthetic arrays (SAR acquisitions).
The target in $\mathbf{x}_t$ is now observed over an angular range $\Delta \psi = \psi^+-\psi^-$, and the observable wavenumbers cover an entire 2D region
\begin{equation}\label{eq:wavenumber_covered_region_aperture}
    \mathcal{K}_{B}(\mathcal{A}) = \bigcup_{\mathbf{x}\in\mathcal{A}} \mathcal{K}_{B}(\mathbf{x})
\end{equation}
whose area is $(4\pi/c)^2 f_0 B \Delta \psi$, located on a circle of radius $4\pi f_0/c $. The resulting radio-sensing image has an additional angular resolution that can be approximated for $A\ll R$ (aperture much less than the range) as
\begin{equation}\label{eq:angular_resolution}
    \rho_\psi \approx \frac{c}{2 f_0 A \sin\psi}.
\end{equation}
The cross-range resolution at distance $R$ is, therefore
\begin{equation}\label{eq:cross_range}
    \rho_{xr}(R) \approx R\, \rho_\psi.
\end{equation}
%

\subsection{Combination of monostatic acquisitions}\label{subsect:2mono_wavenumber_region}

Straightforward ways to improve the overall resolution at the single sensor are: \textit{(i)} increasing the bandwidth $B$, usually not possible given a specific sensing technology and \textit{(ii)} increasing the angular diversity (thus the aperture $A$). In this latter case, massive multiple-input multiple-output (MIMO) radars require fully-digital hardware architecture increasing the cost of the sensor, while increasing the synthetic aperture in a SAR leads to practical challenges \cite{Manzoni2022_TITS}. The only alternative to improve imaging resolution, and also to extend the sensing range, is to use multiple sensors, either belonging to the same physical terminal or to different physical terminals, to acquire a \textit{global} knowledge. A na\"{i}ve approach consists of observing the target from different points of view. This could be representative of a situation in which multiple sensors acquire a radio image independently.

Let us consider two sensors performing monostatic acquisitions with bandwidth $B$ over apertures $\mathcal{A}_1$ and $\mathcal{A}_2$. The overall covered region in the wavenumber domain for the two independent monostatic acquisitions is the union of the two, but depends on whether the two generated images are \textit{incoherently} or \textit{coherently} combined. 

\begin{figure}[!t]
    \centering
    \subfloat[]{\includegraphics[width=\columnwidth]{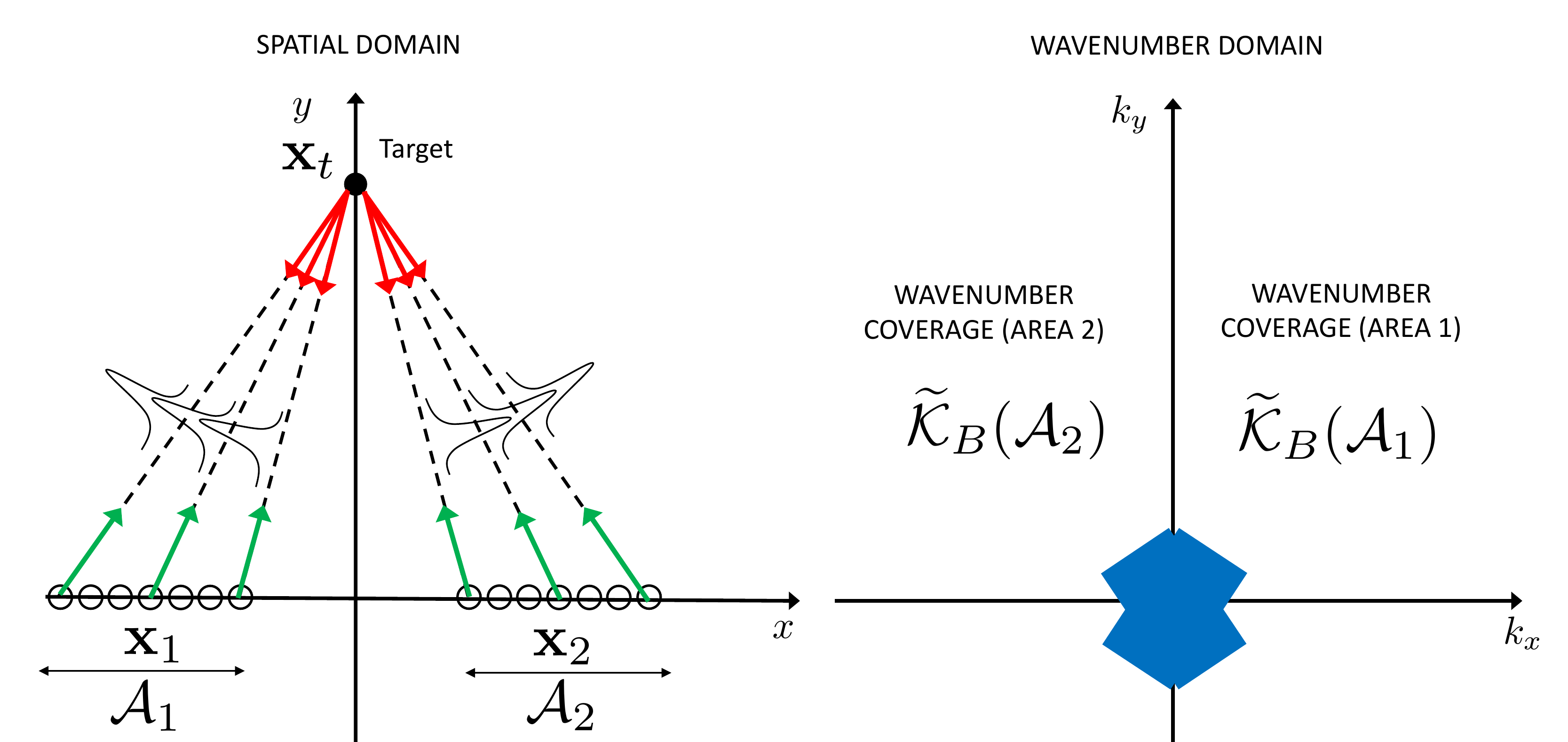}\label{subfig:w6_incoherent}}\\
    \subfloat[]{\includegraphics[width=\columnwidth]{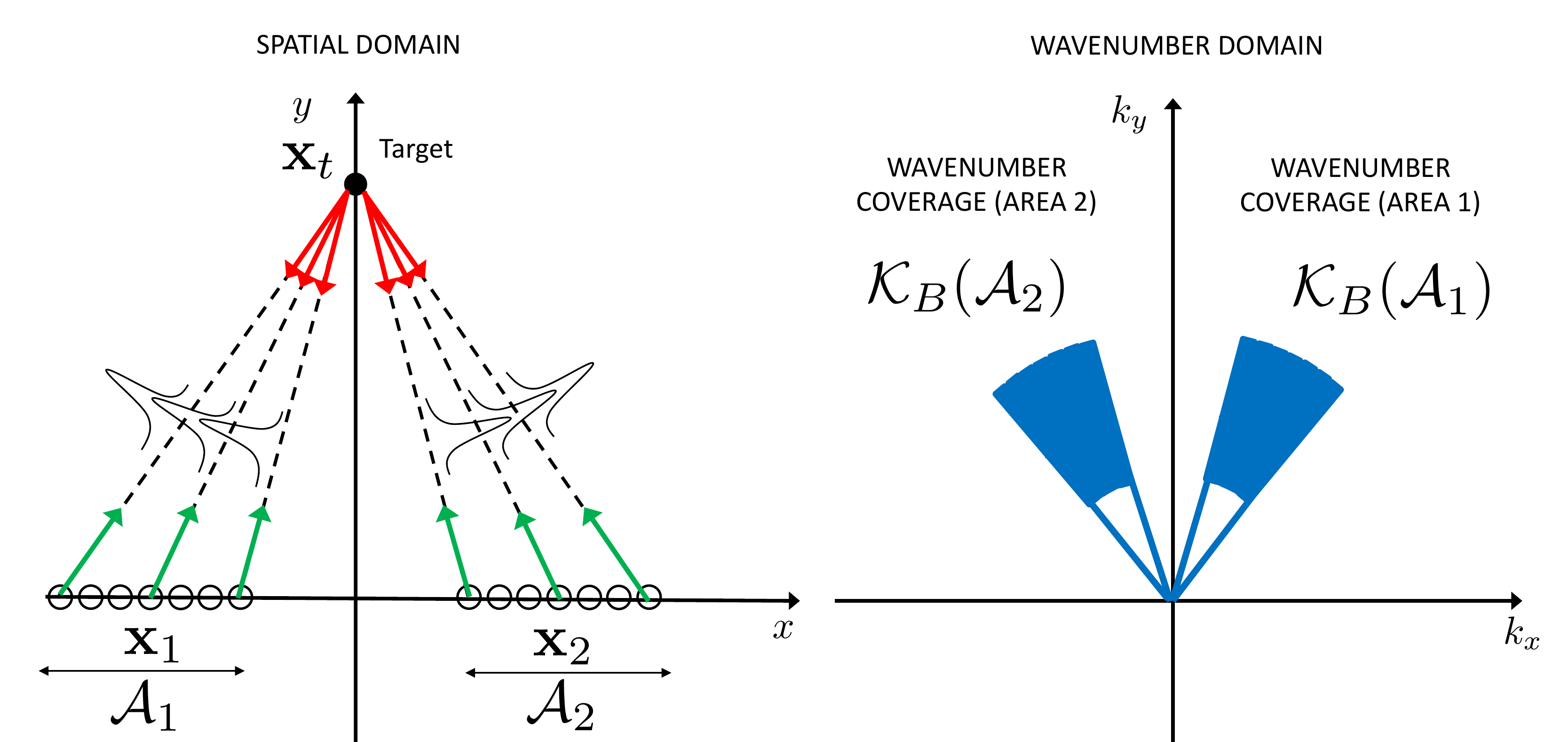}\label{subfig:w6}}\\
    \subfloat[]{\includegraphics[width=\columnwidth]{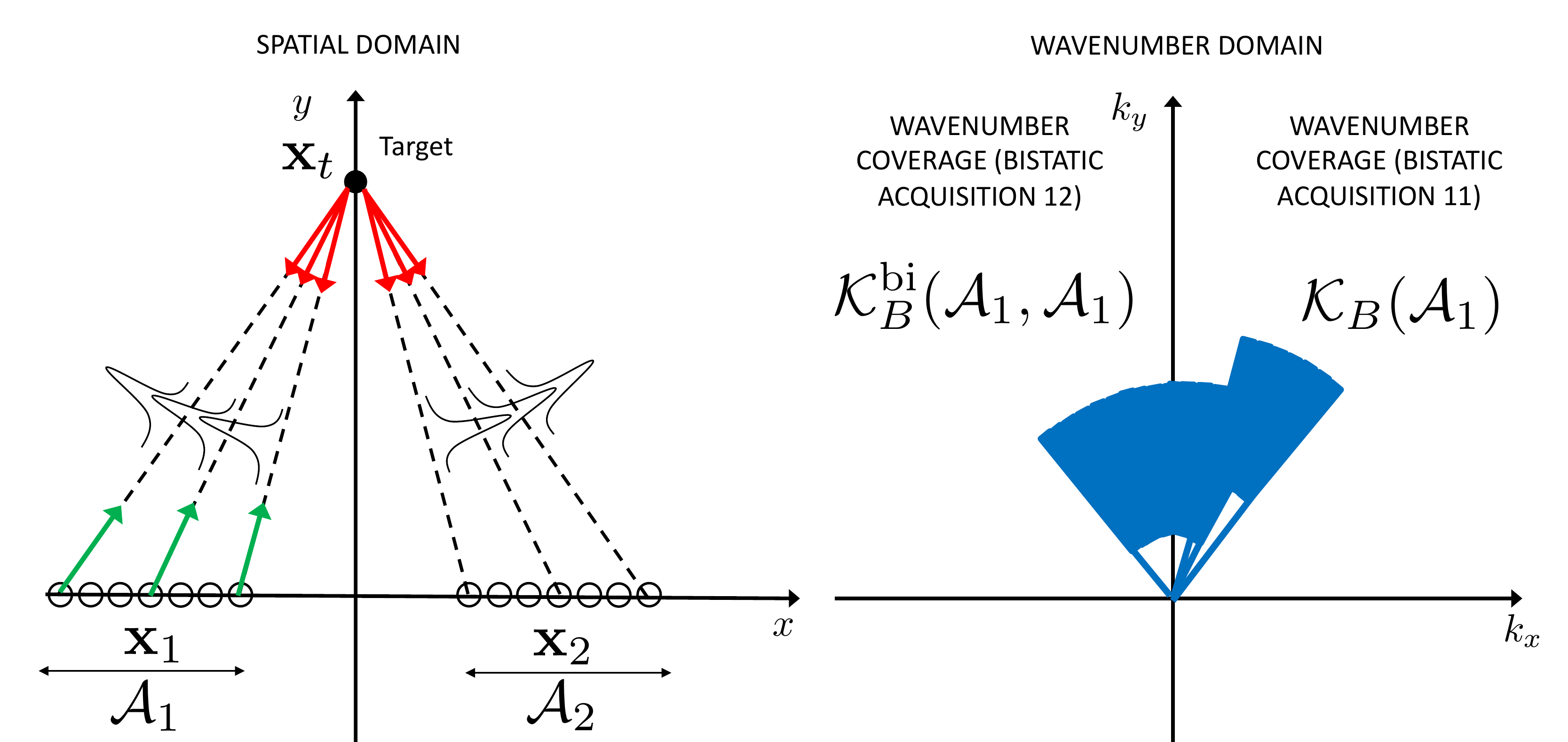}\label{subfig:w7}}
    \caption{ Solutions to increase the image resolution for fixed bandwidth $B$: (\ref{subfig:w6}) incoherent combination of monostatic acquisitions; (\ref{subfig:w6_incoherent}) coherent combination of monostatic acquisitions; (\ref{subfig:w7}) coherent combination of bistatic acquisitions. }
    \label{fig:wavenumbers_bandwidth_bistatic}
\end{figure}

\textbf{Incoherent combination}: The final image is obtained as the average of the absolute value of each single image. The wavenumber coverage spans the four quadrants as shown in Fig. \ref{subfig:w6_incoherent}, and the wavenumber coverage is
\begin{equation}\label{eq:monostatic_comb_coverage_incoherent}
    \widetilde{\mathcal{K}}^{\small\text{mono}}_{B}(\mathcal{A}_1,\mathcal{A}_2) = \widetilde{\mathcal{K}}_{B}(\mathcal{A}_1) \cup \widetilde{\mathcal{K}}_{B}(\mathcal{A}_2),
\end{equation}
where $\widetilde{\mathcal{K}}_{B}(\mathcal{A}_1)$ and $\widetilde{\mathcal{K}}_{B}(\mathcal{A}_2)$ denote \textit{base-band} wavenumber tiles, whose definition follows from \eqref{eq:wavenumber_covered_region_aperture} by considering base-band Tx and Rx wavevectors
\begin{align}
    \widetilde{\mathbf{k}}_{\mathrm{Tx,min}} & = \frac{2\pi  f_\mathrm{min}}{c} [\cos\psi_{\mathrm{Tx}}, \sin\psi_{\mathrm{Tx}}]^\mathrm{T} \label{eq:wavenumber_min_Tx_baseband}\\
    \widetilde{\mathbf{k}}_{\mathrm{Tx,max}} & = \frac{2\pi f_\mathrm{max}}{c} [\cos\psi_{\mathrm{Tx}}, \sin\psi_{\mathrm{Tx}}]^\mathrm{T} \label{eq:wavenumber_max_Tx_baseband}\\
    \widetilde{\mathbf{k}}_{\mathrm{Rx,min}} & = -\frac{2\pi  f_\mathrm{min}}{c} [\cos\psi_{\mathrm{Rx}}, \sin\psi_{\mathrm{Rx}}]^\mathrm{T} \label{eq:wavenumber_min_Rx_baseband}\\
    \widetilde{\mathbf{k}}_{\mathrm{Rx,max}} & = -\frac{2\pi f_\mathrm{max}}{c} [\cos\psi_{\mathrm{Rx}}, \sin\psi_{\mathrm{Rx}}]^\mathrm{T} .\label{eq:wavenumber_max_Rx_baseband}
\end{align}
The single tiles are overlapped in the base-band region; therefore, the overall wavenumber coverage is not significantly improved compared to the single sensor case, but the signal-to-noise ratio is enhanced, as discussed in Section \ref{sec:options_for_cooperation}.

\textbf{Coherent combination}: The final image is obtained by averaging the two single complex-valued images. The wavenumber coverage is
\begin{equation}\label{eq:monostatic_comb_coverage_coherent}
    \mathcal{K}^{\small\text{mono}}_{B}(\mathcal{A}_1,\mathcal{A}_2) = \mathcal{K}_{B}(\mathcal{A}_1) \cup \mathcal{K}_{B}(\mathcal{A}_2),
\end{equation}
as shown in Fig. \ref{subfig:w6}. The spectral content of the experiment is larger compared to single experiments, i.e., the resolution is enhanced, but the two regions are disjoint. This leads to a limited image quality improvement as high grating lobes appear around the location of the target. This might be acceptable for a simple scene with few well-separated targets \cite{Tagliaferri2021_CSAR}, it could prevent the correct imaging when the scene is rich of closely spaced targets of different shapes and dimensions.

\subsection{Multistatic acquisition ($\mathbf{x}_\mathrm{Tx}\neq\mathbf{x}_\mathrm{Rx}$)}\label{subsect:multistatic_wavenumber_region}

%
%


A leap forward for increasing the image quality is to consider generic multistatic configurations, as depicted in Fig. \ref{subfig:w7}, allowing to explore different wavenumber regions, diminishing the grating lobes while preserving the overall resolution. The covered region for a bistatic acquisition where sensor $1$ is Tx and sensor $2$ is Rx is:
\begin{align}
    \mathcal{K}^{\small\text{bi}}_{B}(\mathcal{A}_1,\mathcal{A}_2) = \bigcup_{\substack{\mathbf{x}_1\in\mathcal{A}_1\\\mathbf{x}_2\in\mathcal{A}_2}} \mathcal{K}_{B}(\mathbf{x}_1,\mathbf{x}_2)\label{eq:bistatic_MIMO_wavenumber_region_1}
\end{align}
where $\mathcal{K}_{B}(\mathbf{x}_1,\mathbf{x}_2)$ is the segment in the wavenumber domain obtained from \eqref{eq:wavenumber_region_segment} for two arbitrary Tx and Rx positions. Adding the two monostatic acquisitions, we obtain $\mathcal{K}_{B} = \mathcal{K}^{\small\text{mono}}_{B}(\mathcal{A}_1,\mathcal{A}_2) \,\cup \,\mathcal{K}^{\small\text{bi}}_{B}(\mathcal{A}_1,\mathcal{A}_2)$. As shown in Section \ref{sec:orchestration}, the resulting image shows a better quality in terms of both resolution and grating lobes level. The next step is to devise a general image formation algorithm to synthesize the image of the target for any sensing network topology and capabilities of the involved terminals (bandwidth, aperture, etc.).

\subsection{Resolution of the image}\label{subsect:resolution}

The image resolution in Cartesian coordinates can be quantified by evaluating the convex hull of the wavenumber coverage of the acquisition $\mathcal{K}$, e.g., \eqref{eq:monostatic_comb_coverage_coherent} or \eqref{eq:bistatic_MIMO_wavenumber_region_1}, namely:
 \begin{equation}\label{eq:convhull}
     \mathcal{K}_\mathrm{hull} = \left\{\sum_{i=1}^{\lvert\mathcal{K}\rvert} \alpha_i \mathbf{k}_i \,\bigg\lvert\, \mathbf{k}_i\in \mathcal{K}, \alpha_i\geq 0,\, \sum_{i}\alpha_i=1\right\}
 \end{equation}
where $\lvert\mathcal{K}\rvert$ is the cardinality of $\mathcal{K}$. Eq. \eqref{eq:convhull} can be generalized to arbitrary sets with an uncountable number of elements with proper adjustments. From $\mathcal{K}_\mathrm{hull}$ the resulting image resolution along $x$ and $y$ can be approximated as
\begin{equation}\label{eq:2Dresolution}
     \rho_x\approx \frac{2\pi}{\Delta k_x},\,\,\,\rho_y\approx \frac{2\pi}{\Delta k_y},
\end{equation}
where 
\begin{align}
    \Delta k_x & = \mathrm{sup}\left\{\big\lvert[\mathbf{k}_i]_x-[\mathbf{k}_j]_x\big\rvert \, , \, \mathbf{k}_i,\mathbf{k}_j\in \mathcal{K}_\mathrm{hull}\right\}\\
    \Delta k_y & = \mathrm{sup}\left\{\big\lvert[\mathbf{k}_i]_y-[\mathbf{k}_j]_y\big\rvert \, , \, \mathbf{k}_i,\mathbf{k}_j\in \mathcal{K}_\mathrm{hull}\right\}
\end{align}
denote the width of the image spectral coverage along $x$ and $y$. 

\section{System Model and Image Formation}
\label{sec:focusing_algorithms}

This section introduces the system model and the radio image formation by back-projection, derived from the wavenumber domain. We consider to have a sensing network made by $L$ terminals in the environment, whose phase center is located in $\mathbf{x}_\ell$, $\ell=1,...L$. Each of the terminals is equipped with $N$ Tx elements and $M$ Rx elements. The position of the $n$-th Tx element of the $\ell$-th terminal is denoted by $\mathbf{x}_{\mathrm{T},n}^\ell$, while the $m$-th Rx element of the $k$-th terminal is $\mathbf{x}_{\mathrm{R},m}^k$.  Each terminal can be either static or in motion, thus the Tx and Rx elements can be displaced in space (as in a MIMO array) or in time-space (as in a synthetic array). The present treatment is general and applies to both cases.

The goal of the $L$ terminals is to image a point target located in $\mathbf{x}$. According to the specific cooperation policy, discussed in Section \ref{sec:options_for_cooperation}, we can define an association matrix $\mathbf{B}\in\mathbb{B}^{L\times L}$ that pairs the allowed monostatic or bistatic sensing Tx-Rx pairs. The $\ell k$-th entry of $\mathbf{B}$ is as follows:
\begin{equation}
    [\mathbf{B}]_{\ell k} = \begin{dcases}
    1 & \text{if $\ell$-th Tx and $k$-th Rx are paired}\\
    0 & \text{otherwise}.
\end{dcases}
\end{equation}
Thus, the possible number of multistatic pairs and measurement channels is, respectively, $\| \mathbf{B}\|_0$ and $\| \mathbf{B}\|_0 NM$, respectively. The $\ell$-th Tx terminal emits the modulated signal around carrier frequency $f_0$:
\begin{equation}
    s_\ell(t) = g_\ell(t)e^{j2\pi f_0 t}
\end{equation}
where $g_\ell(t)$ is the complex-valued base-band pulse of bandwidth $B$. Notice that each Tx-Rx pair works on the same spectrum portion, thus the base-band pulses shall be designed to achieve some orthogonality conditions (e.g., orthogonal in time or codes \cite{deWit2011}). 

The generic model of the Rx signal at the $k$-th terminal, $m$-th Rx element, due to the transmission from the $n$-th Tx element of the $\ell$-th terminal (after demodulation) is:
\begin{equation}\label{eq:rx_signal}
\begin{split}
    y_{nm}^{\ell k}(t) = [\mathbf{B}]_{\ell k}\; \beta_{\ell k} \, & g_\ell(t-\tau_{nm}^{\ell k}-\Delta t_{\ell k}) \times \\
    & \times e^{-j2\pi f_0 (\tau_{nm}^{\ell k}+\Delta t_{\ell k})} + z_{nm}^{\ell k}(t)
\end{split}
\end{equation}
where \textit{(i)} $\beta_{\ell k}$ is the complex scattering amplitude accounting for both the two-way path-loss to-from the target as well as the radar cross section (RCS) of the target \cite{skolnik}, \textit{(ii)} $\tau_{nm}^{\ell k}$ is the two-way propagation delay, equal to
\begin{equation}\label{eq:delay}
    \tau_{nm}^{\ell k} = \frac{\big\|\mathbf{x}-\mathbf{x}_{\mathrm{T},n}^\ell \big\|}{c} + \frac{\big\|\mathbf{x}_{\mathrm{R},m}^k-\mathbf{x}\big\|}{c},
\end{equation}
\textit{(iii)} $\Delta t_{\ell,k}$ models any residual clock synchronization error between the $\ell$-th and the $k$-th terminals and \textit{(iv)} $z_{nm}^{\ell k}(t)\in\mathcal{CN}(0, \sigma_z^2 \delta(t)\delta_{\ell-k}\delta_{n-m})$ is the additive white noise, uncorrelated across different channels. 
\begin{figure}[!t]
    \centering
    \includegraphics[width=0.6\columnwidth]{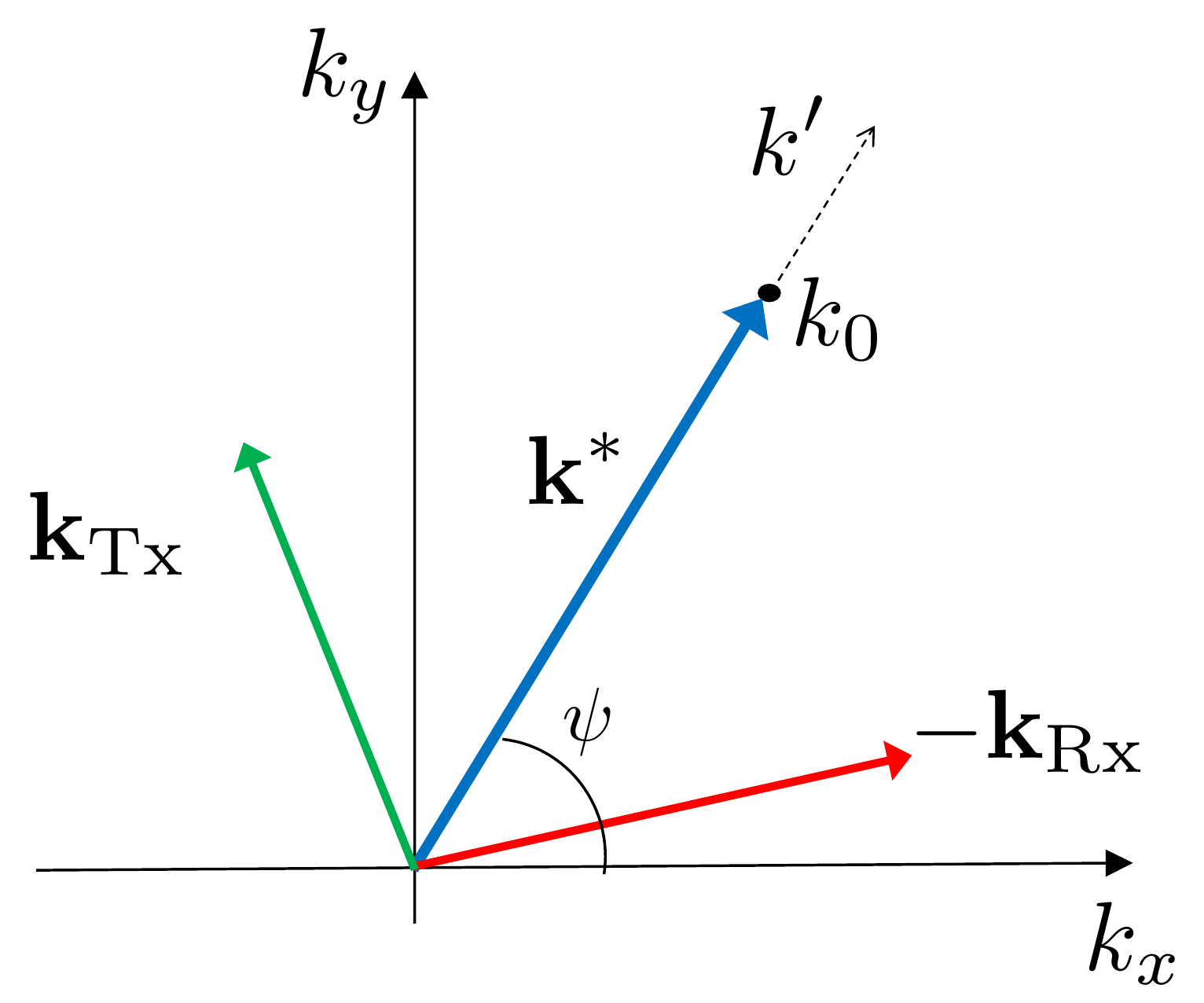}
    \caption{Pictorial representation of wavevectors, angles, and quantities playing a role in deriving the back-projection integral starting from the wavenumber domain.}
    \label{fig:BP_plot_wavenumbers}
\end{figure}
%
\subsection{From Wavenumbers to the Back-Projection} 
Let us assume the perfect synchronization between the intended Tx and Rx terminals. The image formation from measured data \eqref{eq:rx_signal} shall be general enough to take into account any possible Tx-Rx position and configuration in space (near-field vs. far-field, arbitrary orientation of the terminals, monostatic vs. bistatic acquisition) as well as any possible employed bandwidth (narrowband vs. wideband target response across the array). Here we derive a general image formation algorithm starting from the wavenumber-covered region.

Recalling \eqref{eq:Rxsignal_FT} in Section \ref{sect:FEDT}, we can write the inverse Fourier transform to obtain the reflectivity of the scene in global coordinates as follows:
\begin{equation}\label{eq:inverse_FT}
    \gamma(\mathbf{x}) = \int_{-\infty}^{+\infty} \int_{-\infty}^{+\infty}  \Gamma(\mathbf{k})\,e^{j\mathbf{k}^T \mathbf{x}} \frac{d\mathbf{k}}{(2\pi)^2}
\end{equation}
where $\mathbf{k}=[k_x,k_y]^T$ are the excited wavenumbers in global coordinates, obtained by \eqref{eq:enlight_wavenumber} in Section \ref{sect:FEDT}, and the integral spans all the wavenumber domain. In practice, the double integral is limited to the spectral coverage of the given radio sensing experiment. Let us focus on a generic bistatic radio sensing experiment, as the one described in Section \ref{subsect:multistatic_wavenumber_region}, where the Tx positions $\mathbf{x}_\mathrm{Tx}$ span aperture $\mathcal{A}_\mathrm{Tx}$ and Rx positions $\mathbf{x}_\mathrm{Rx}$ span aperture $\mathcal{A}_\mathrm{Rx}$. The wavenumber covered region is therefore $\mathcal{K}^\text{bi}_B(\mathcal{A}_\mathrm{Tx}, \mathcal{A}_\mathrm{Rx})$, defined as in \eqref{eq:bistatic_MIMO_wavenumber_region_1}. It is now interesting to operate a change of variable within the integral \eqref{eq:inverse_FT} in order to obtain a separable integration domain and split the double integral into two cascaded integrals. In particular, we can express the generic wavenumbers $\mathbf{k}=[k_x,k_y]^T$ as function of a scalar wavenumber $k$ and an \textit{observation angle} $\psi$:
\begin{align}
    k_x &= k \sin{\psi} = (k_0+k')\sin{\psi} \\
    k_y &= k \cos\psi = (k_0+k')\cos{\psi} 
\end{align}
where $k_0$ is the central wavenumber (of the carrier frequency) and $k'$ is the baseband wavenumber, pertaining to the bandwidth of the Tx pulse $g(t)$. The observation angle is 
\begin{equation}
    \psi = \frac{\psi_\mathrm{Tx} + \psi_\mathrm{Rx}}{2}
\end{equation}
thus pertains to the \textit{virtual monostatic} channel, located halfway between Tx and Rx. Notice that we are now using the equivalent monostatic wave vector $\mathbf{k}^*$ in \eqref{eq:enlight_wavenumber}, which is made by the difference between the Tx wavevector $\mathbf{k}_\mathrm{Tx}$ and the Rx one $\mathbf{k}_\mathrm{Rx}$, as represented in Figure \ref{fig:BP_plot_wavenumbers}.

Assuming that the point target to be imaged is located in the origin of the global coordinates, i.e., $\mathbf{x}_t=[0,0]^T$, the integral \eqref{eq:inverse_FT} becomes, after the change of variable $(k_x,k_y) \rightarrow (k',\psi)$,
\begin{equation}
\label{eq:TDBP_k_psi}
    \gamma(\mathbf{0}) = \int \limits_{\psi_{min}}^{\psi_{max}} \int \limits_{k_{min}'}^{k_{max}'} (k_0+k') \,\Gamma(k_0+k',\psi) \frac{dk' d\psi}{(2\pi)^2}
\end{equation}
where the double integral \eqref{eq:inverse_FT} on the wavenumer domain is now replaced by a first integral over the baseband frequency, spanning the employed bandwidth $B$ through the explicit dependency on $k'$, and a second integral over all the angular positions of the Tx-Rx pairs through the explicit dependency on $\psi$. Notice that the inverse Fourier transform in the new polar coordinate system in the wavenumber domain includes the Jacobian of the transformation $ dk_xdk_y = (k_0+k') dk'd\psi$.
The limits of the integrals can be derived from basic geometrical considerations as follows (see Section \ref{sect:FEDT}):
\begin{equation}
    \begin{split}
        k_{min}' = \frac{2 \pi f_{min}}{c}\sqrt{2 + 2\cos{(\psi_{\mathrm{Tx}}-\psi_{\mathrm{Rx}})}}\\
        k_{max}' = \frac{2 \pi f_{max}}{c}\sqrt{2 + 2\cos{(\psi_{\mathrm{Tx}}-\psi_{\mathrm{Rx}})}}
    \end{split}
\end{equation}
where $\psi_{\mathrm{Tx}}$ and $\psi_{\mathrm{Rx}}$ are the angular positions at which the two operating sensors observe the target in the scene. The integral over $\psi$ spans all the Tx-Rx pairs positions, over the apertures, through a suitable angular domain.

We can now plug the Rx signal model (\ref{eq:Rxsignal_FT}) (Section \ref{sect:FEDT}) into (\ref{eq:TDBP_k_psi}), obtaining: 
\begin{align}
\begin{split}
    \label{eq:BP_wavenumbers}
        \gamma(\mathbf{0}) = \int \limits_{k_{min}'}^{k_{max}'} \int \limits_{\psi_{min}}^{\psi_{max}}  &(k_0+k') r(\mathbf{x}_{\mathrm{Tx}},\mathbf{x}_{\mathrm{Rx}}) \times \\
        & \times e^{j (k_0+k') (R_{\mathrm{Tx}}(\mathbf{0}) + R_{\mathrm{Rx}}(\mathbf{0}))}  \frac{dk'd\psi}{(2\pi)^2}.
\end{split}
\end{align}
Notice that \eqref{eq:BP_wavenumbers} is the expression of a \textit{filtered back-projection}: the received signal $r(\mathbf{x}_{\mathrm{Tx}},\mathbf{x}_{\mathrm{Rx}})$ is first multiplied by $(k_0+k')$ (the \textit{filter}), and then by a phase term compensating for the macroscopic bistatic delay between the target and the Tx/Rx positions. The result is integrated over the base-band frequencies $k'$ and over the Tx and Rx apertures (through the virtual monostatic angle $\psi$). The back-projection follows two steps:

\textbf{Range compression:} The Rx signal $r(\mathbf{x}_{\mathrm{Tx}},\mathbf{x}_{\mathrm{Rx}})$ in \eqref{eq:BP_wavenumbers} is linked to the one in \eqref{eq:rx_signal} after the so-called range compression, i.e., the integral over $k'$: 
\begin{equation}
\label{eq:range_compression}
y(\mathbf{x}_{\mathrm{Tx}},\mathbf{x}_{\mathrm{Rx}}) \hspace{-0.1cm}= \hspace{-0.1cm}\int \limits_{k_{min}'}^{k_{max}'} \hspace{-0.1cm}r(\mathbf{x}_{\mathrm{Tx}},\mathbf{x}_{\mathrm{Rx}})e^{j k' (R_{\mathrm{Tx}}(\mathbf{0}) + R_{\mathrm{Rx}}(\mathbf{0}))} \frac{dk'}{2\pi}
\end{equation}
Although omitted, the result of \eqref{eq:range_compression} is a function of time. 

\textbf{Focusing:} The integral over $\psi$ is referred to as focusing:
\begin{equation}
\label{eq:bp_final}
    \gamma(\mathbf{0}) = \int \limits_{\psi_{min}}^{\psi_{max}} y(\mathbf{x}_{\mathrm{Tx}},\mathbf{x}_{\mathrm{Rx}})e^{j k_0 (R_{\mathrm{Tx}}(\mathbf{0}) + R_{\mathrm{Rx}}(\mathbf{0}))} \frac{d\psi}{2\pi}
\end{equation}
where we neglect the \textit{filter} $(k_0+k')$, an approximation that holds for most passband radio sensing experiments, where $f_0\gg B$. It can be demonstrated that the back-projection integral \eqref{eq:BP_wavenumbers} applies to any geometry and physical configuration of Tx and Rx, and it degenerates to the well-known DFT approach for far-field acquisitions, (i.e, with plane waves, regular array geometries narrow bandwidth, $f_0\gg B$). Extension to generic targets not located in the origin of the coordinate system follows from the definition \eqref{eq:inverse_FT}, with additional path-induced phase terms in the back-projection integral \eqref{eq:BP_wavenumbers}.  
\begin{algorithm}[t!]
\caption{Back-Projection algorithm}\label{al:BP}
\begin{algorithmic}[1]
\ENSURE Image $I_{\ell k}(\overline{\mathbf{x}})$ for all $\overline{\mathbf{x}}$ in the grid
\STATE \textbf{Initialize:}
\STATE $I_{\ell k}(\overline{\mathbf{x}}) \gets 0$ for all $\overline{\mathbf{x}}$ in the grid

\STATE \textit{\%for all the image pixels}
\FOR{$\overline{\mathbf{x}}$ in the grid}
\STATE \textit{\%for all the Tx}
\FOR{$n=1$ \textbf{to} $N$}
\STATE \textit{\%for all the Rx}
\FOR{$m=1$ \textbf{to} $M$}
\STATE $I_{nm}^{\ell k}(\overline{\mathbf{x}}) \gets 0$
\STATE (1) Extrapolate the signal value at the time instant $\overline{\tau}_{nm}^{\ell k}$, corresponding to pixel $\overline{\mathbf{x}}$: $I_{nm}^{\ell k}(\overline{\mathbf{x}}) \gets y_{nm}^{\ell k} (t=\overline{\tau}_{nm}^{\ell k})$
\STATE (2) Phase rotation by: $I_{nm}^{\ell k}(\overline{\mathbf{x}}) \gets I_{nm}^{\ell k}(\overline{\mathbf{x}}) e^{j2\pi f_0\overline{\tau}_{nm}^{\ell k} }$ 
\STATE (3) Accumulate: $I_{\ell k}(\overline{\mathbf{x}}) \gets I_{\ell k}(\overline{\mathbf{x}}) + I_{nm}^{\ell k}(\overline{\mathbf{x}})$
\ENDFOR
\ENDFOR
\ENDFOR








\end{algorithmic}
\end{algorithm}

\subsection{Practical Implementation of the Back-Projection}
\label{sec:TDBP_section}
The back-projection \eqref{eq:BP_wavenumbers} can be adapted to the system model in \eqref{eq:rx_signal} with straightforward modifications.  
Let us consider the $\ell k$-th generic bistatic Tx-Rx pair. The image is defined as the spatial map of the complex reflectivity of the scene, evaluated on a pre-defined pixel grid. For the pixel in position $\overline{\mathbf{x}}$, the image is formed as follows:
\begin{equation}\label{eq:BP}
    I_{\ell k}(\overline{\mathbf{x}}) = \sum_{n} \sum_{m} y_{nm}^{\ell k} (t=\overline{\tau}_{nm}^{\ell k}) e^{j 2 \pi f_0 \overline{\tau}_{nm}^{\ell k}}
\end{equation}
where
\begin{equation}\label{eq:delay_pixel}
    \overline{\tau}_{nm}^{\ell k} = \frac{\big\|\overline{\mathbf{x}}-\mathbf{x}_{\mathrm{T},n}^\ell \big\|}{c} + \frac{\big\|\mathbf{x}_{\mathrm{R},m}^k-\overline{\mathbf{x}}\big\|}{c}
\end{equation}
is the bistatic delay corresponding to the pixel position.
It is interesting to note that (\ref{eq:BP}) is the discrete implementation of \eqref{eq:bp_final} and it implies three specific operations:
\begin{enumerate}
    \item interpolation of the Rx data in the temporal position corresponding to the desired sample, represented by $t=\overline{\tau}_{nm}^{\ell k}$. Since the Rx data is sampled, this operation is necessary to \textit{migrate} the signal acquired by the $nm$-th channel in the correct pixel location of the resulting sensing image;
    \item phase rotation by $e^{j 2 \pi f_0 \overline{\tau}_{nm}^{\ell k}}$, allowing to account for the exact law of variation of the wave phase across different channels, which includes the wave-front curvature as well as any possible effect that arises when array geometry is not straight or uniform;
    \item accumulation of the interpolated and phase-rotated data over all channels.
\end{enumerate}
The back-projection routine is summarized by Algorithm \ref{al:BP}. It is worth underlining that the back-projection requires both the knowledge of the position of the Tx and Rx measurement channels, i.e., $\mathbf{x}_{\mathrm{T},n}^\ell$ and $\mathbf{x}_{\mathrm{R},m}^k$, and the clock and phase synchronization of the intended Tx and Rx terminals. An error in the absolute position of the aperture will lead to the wrong localization of targets, while an error in the relative position between sensors can lead to a strong defocusing of the image. In turn, this condition can cause missed detections of targets. Knowing the precise position of each sensor is usually possible in sensing networks where the terminals are static, e.g., BS or RSU, while for mobile sensing networks, e.g., vehicular networks, other advanced techniques apply \cite{Manzoni2022_TITS}.
The effects of a wrong time/phase synchronization between terminals are very similar, leading to mis-localization and a remarkable defocusing of the entire scene. 

\section{Options for Cooperation}
\label{sec:options_for_cooperation}
This section outlines and discusses different practical options for cooperation in networked sensing scenarios. We focus on \textit{how} the image of the scene is generated from the single ones, the latter obtained with the back-projection or one of its low-complexity variants (Section \ref{sec:TDBP_section}). The fundamental question in this framework is how much gain in spatial resolution we can achieve by enforcing cooperation between terminals, and which is the cost and related pitfalls. The options for cooperation are the following:

\begin{figure}[!t]
    \centering
    \includegraphics[width=0.9\columnwidth]{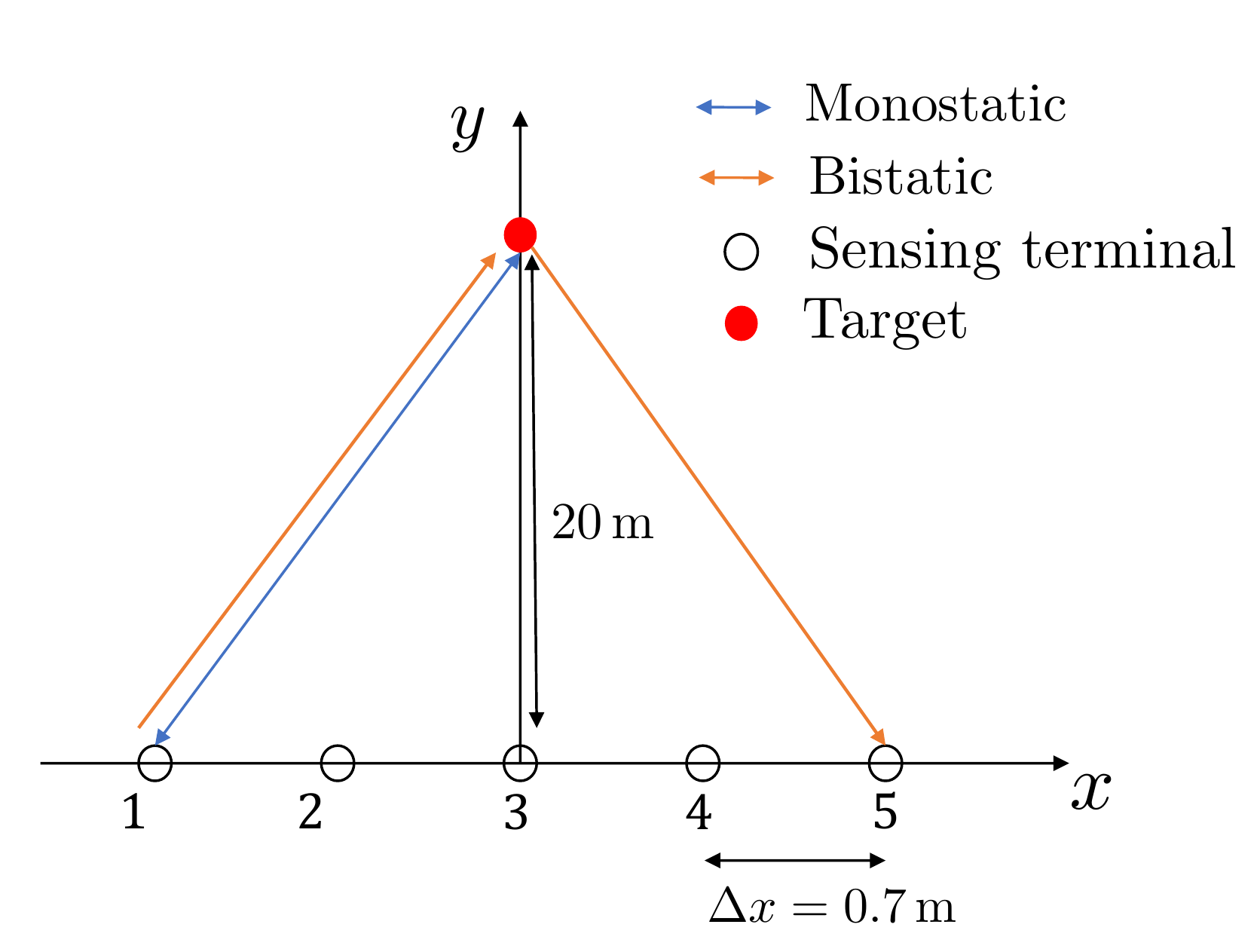}
    \caption{Considered sensing network topology (image not in scale).}
    \label{fig:geometry_1}
\end{figure}

\subsubsection{Fusion of monostatic images} Each terminal produces its own image of the environment $I_{\ell \ell}(\mathbf{x})$ and share it with the others using a suitable communication link (or set of communication links). 

\subsubsection{Fusion of multistatic images} Each terminal produces its own image of the environment $I_{\ell \ell}(\mathbf{x})$ as well as it contributes to the formation of bistatic images $I_{\ell k}(\mathbf{x})$. The latter phase implies full cooperation at the acquisition level, as well as cooperation for exchanging the images with other terminals.

\subsubsection{Orchestrated sensing} A subset of the whole set of available sensing terminals contributes to the formation of both monostatic and multistatic images, attempting to achieve the same imaging quality of the full cooperation. Still, the cooperation is enforced in both acquisition and exchange phases.

We look at cooperation from the DTT perspective, providing quantitative insights for each option. To this aim, we consider the reference exemplary scenario in Fig. \ref{fig:geometry_1}. A total of $L=5$ terminals are present, aligned with the $x$ axis and displaced by $\Delta x= $ 0.7 m, each one equipped with $N=1$ Tx antenna and an $M$ Rx antennas, organized in a $\lambda_0/2$ uniform linear array. A single point target is located at $\mathbf{x}_t = [0,20]^T$. The employed bandwidth is $B=500$ MHz, at the carrier frequency of $f_0=28$ GHz. The $M$ Rx antennas are set to guarantee the same spatial resolution along range and cross-range directions, namely $\rho_r = \rho_{xr} = 30$ cm. The selected configuration is only exemplary as the derived concepts are general and can be applied to \textit{any} geometry and physical sensor capabilities.

\begin{figure}[!t]
\centering
\subfloat[][]{\includegraphics[width=0.5\columnwidth]{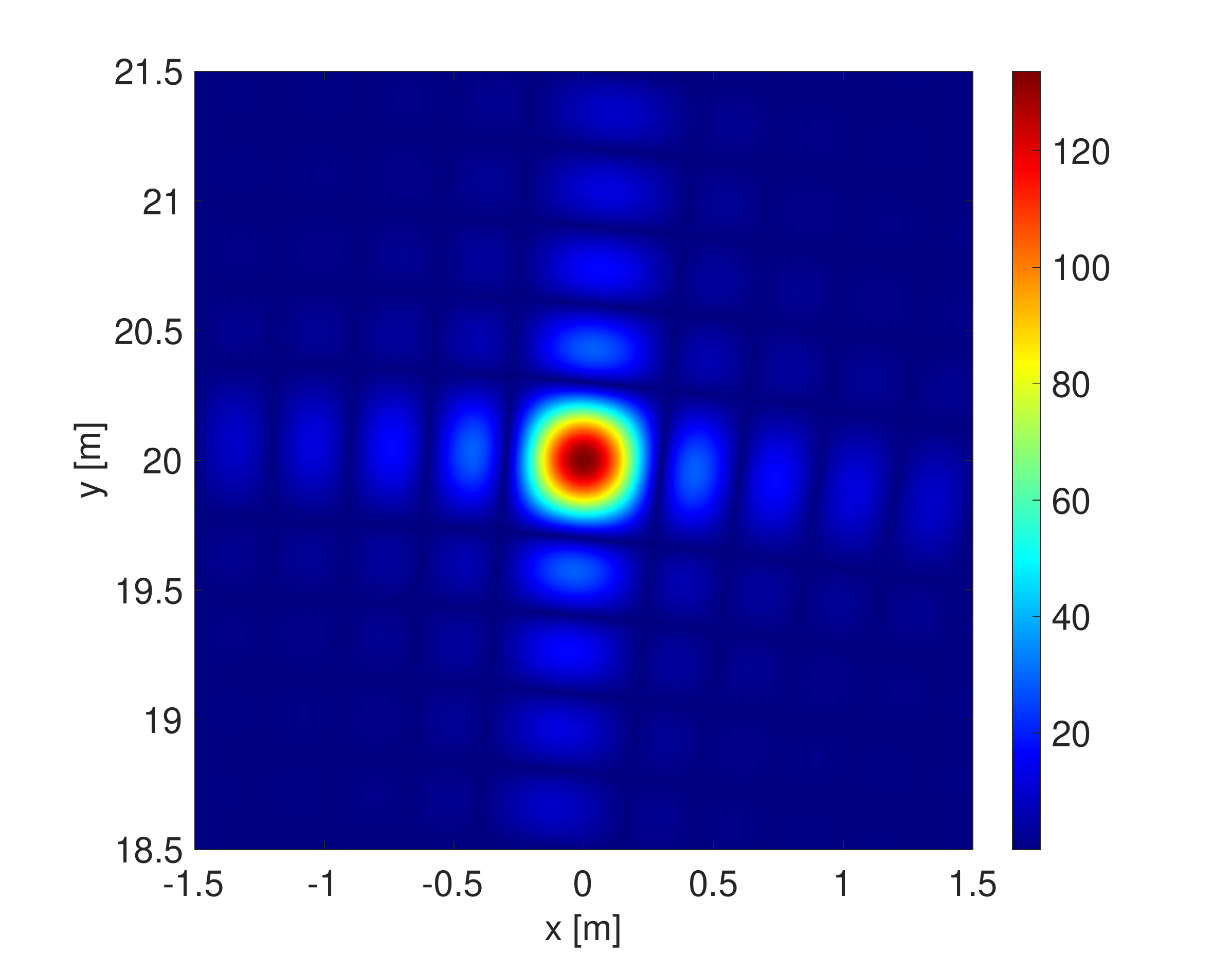}\label{subfig:monostatic-1}}
\subfloat[][]{\includegraphics[width=0.5\columnwidth]{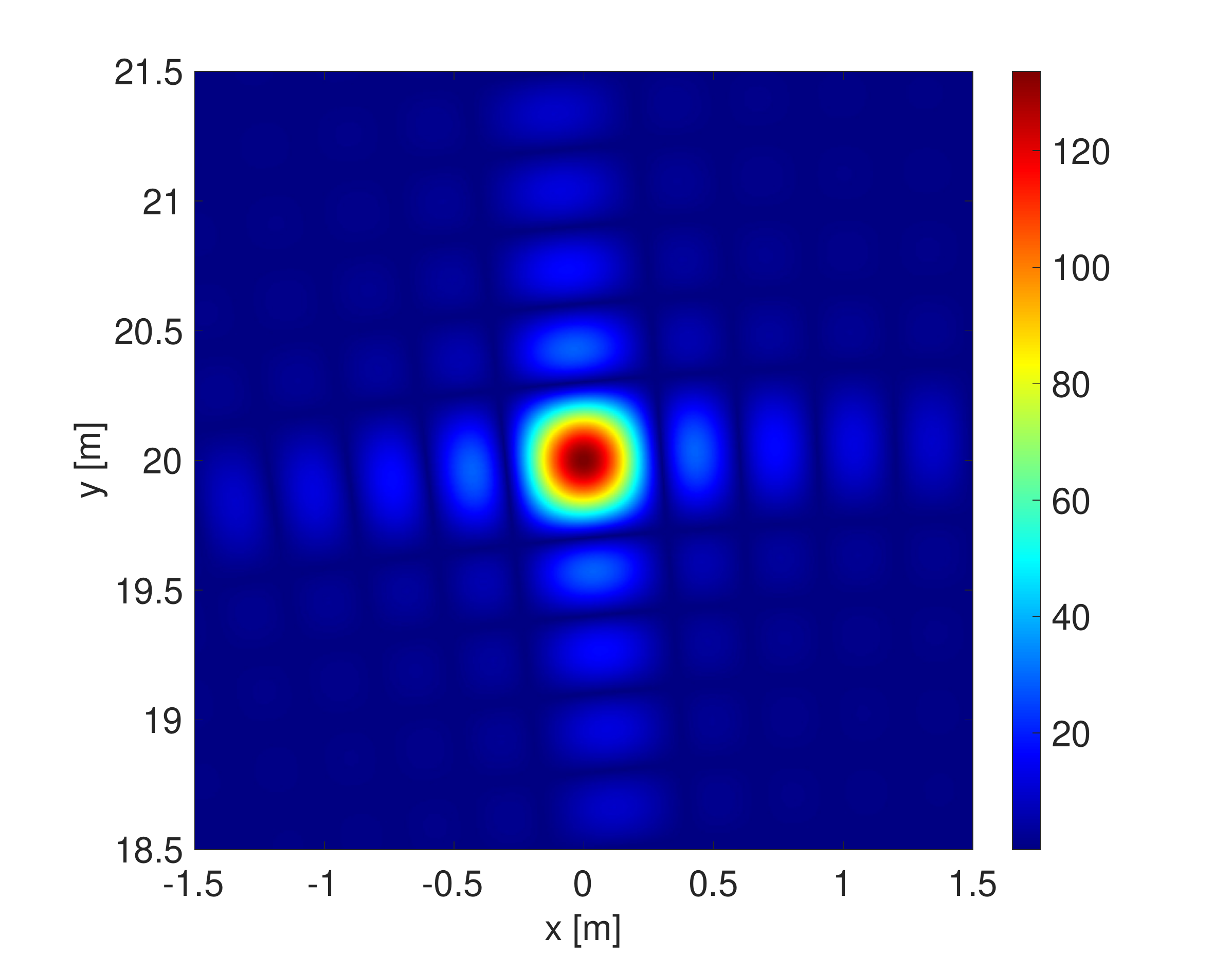}\label{subfig:monostatic-5}}\\
\subfloat[][]{\includegraphics[width=0.5\columnwidth]{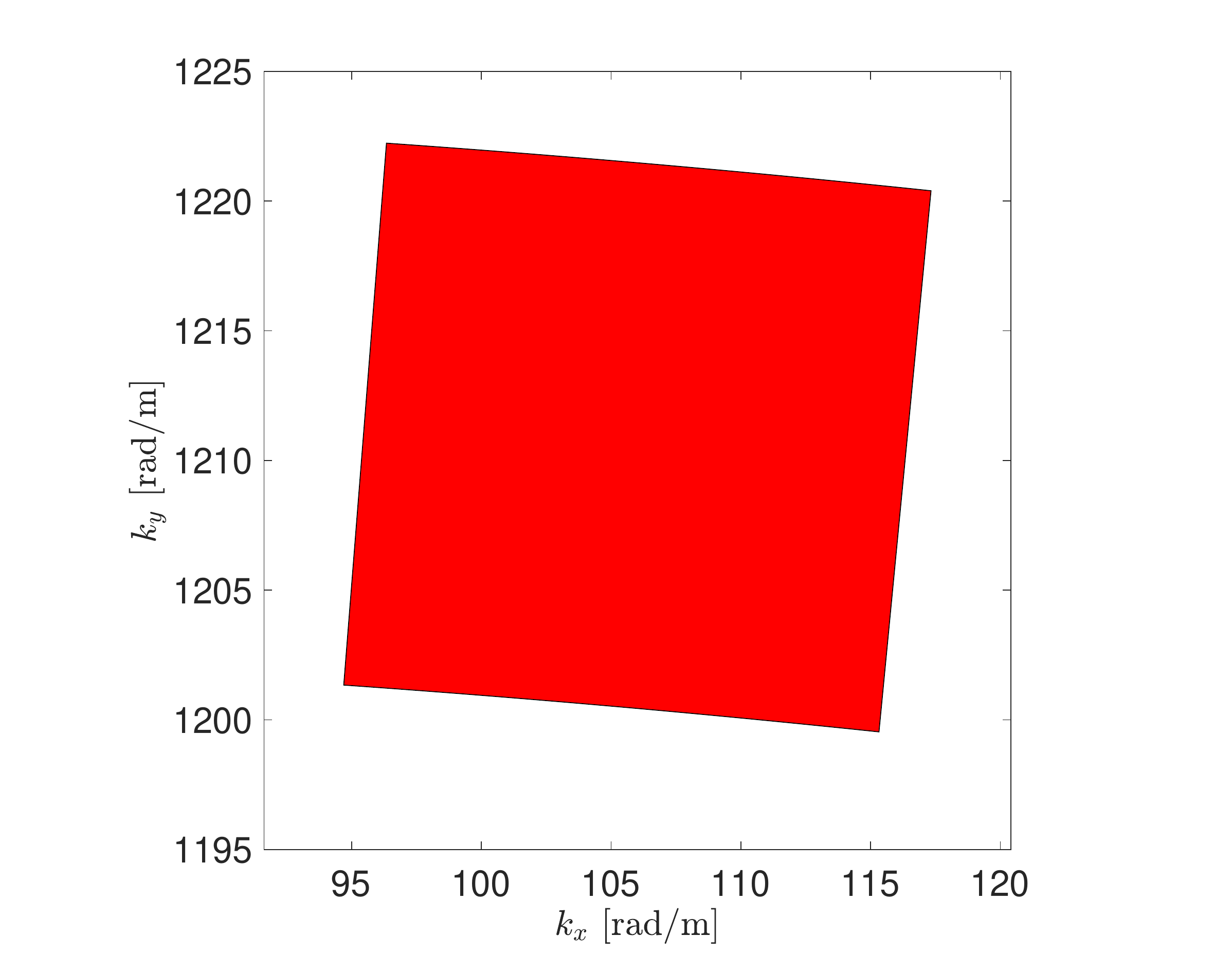}\label{subfig:wavenumber_domain_1}} 
\subfloat[][]{\includegraphics[width=0.5\columnwidth]{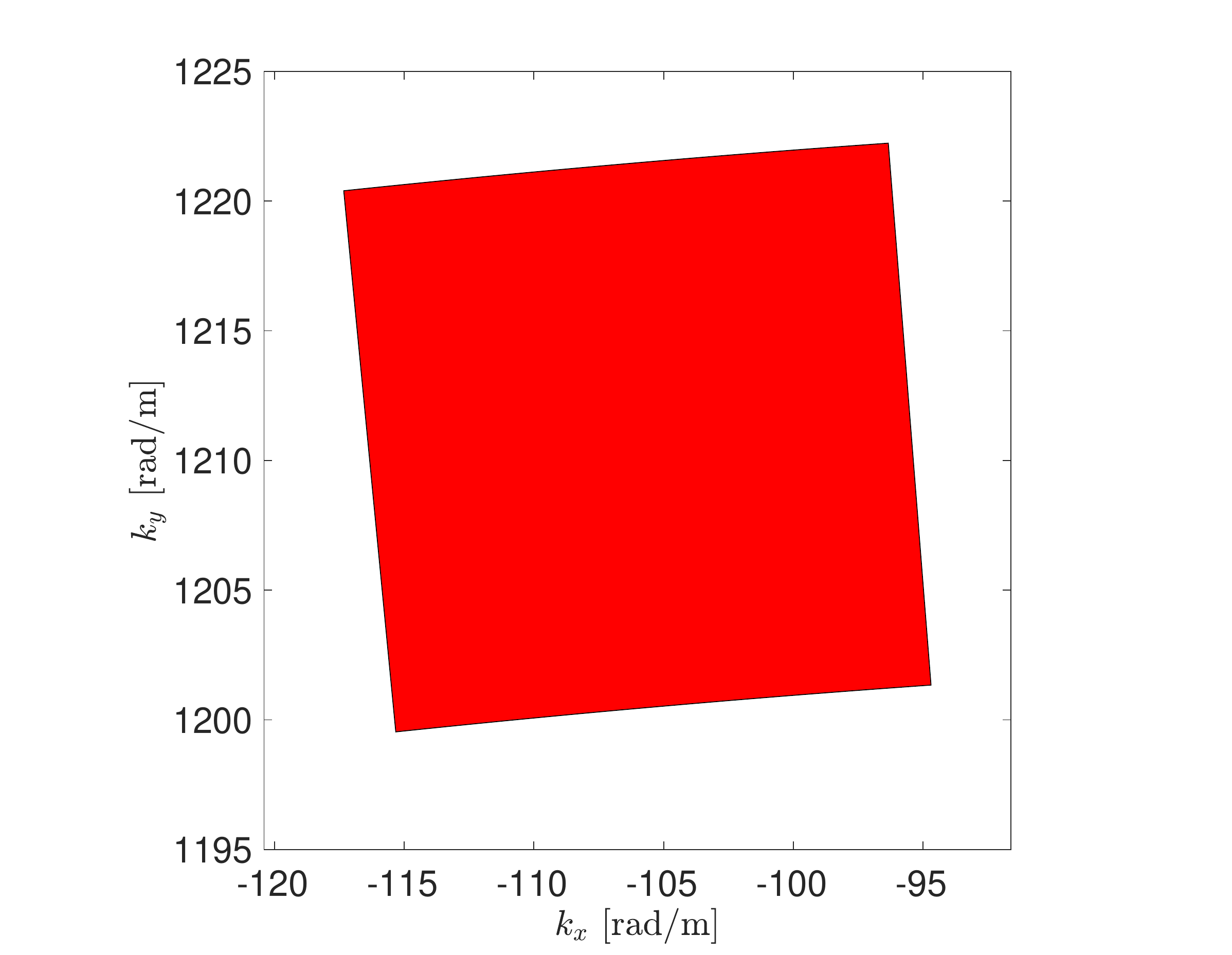}\label{subfig:wavenumber_domain_5}}
\caption{Example of imaging of a point target from different, well-separated, points of view (different sensing terminals). First row (\ref{subfig:monostatic-1},\ref{subfig:monostatic-5}) shows the images, the second row (\ref{subfig:wavenumber_domain_1},\ref{subfig:wavenumber_domain_5}) shows the wavenumber coverage. }
\label{fig:monostatic}
\end{figure}

When the terminals do not exchange any information with the others (no cooperation), the only possibility is to produce monostatic images (e.g., with the back-projection method in \eqref{eq:BP}). This is the simplest case and yields an environment representation that is limited by the available bandwidth and physical aperture at each terminal. No clock and/or phase synchronization is required among terminals. An example of no cooperation is provided in Fig. \ref{fig:monostatic}: Figs. \ref{subfig:monostatic-1} and \ref{subfig:monostatic-5} (first row) show the point target image using two of the sensors in Fig. \ref{fig:geometry_1}, while Figs. \ref{subfig:wavenumber_domain_1} and \ref{subfig:wavenumber_domain_5} (second row) plot the corresponding covered regions in the wavenumber domain. The resolution, i.e, width of the main lobe, in the images is inversely proportional to the spectral coverage, following the simple rule in Section \ref{subsect:resolution}. It is worth noticing that this is the focus of the major part of available literature on ISAC systems, where the focus is the single terminal design and not the potential cooperation \cite{Liu_survey}. In the following, we detail the options for cooperation with relevant benefits/disadvantages.

\subsection{Fusion of Monostatic Images}
\label{sec:monostatic_no_sharing}

The first attempt at cooperation is the exchange of the single images formed at each terminal by monostatic acquisitions. This is the case where there exists a communication network between the sensing terminals that allows the information exchange, or, each terminal is connected to a remote processing unit that gathers, elaborates, and redistributes information from/to all the terminals \cite{demirhan2023cellfree}. Therefore there are two options for cooperation, with increasing complexity:

\subsubsection{Incoherent sum of images}
The easiest method to combine the images is in an incoherent way. The result is:
\begin{equation}\label{eq:incoherent_sum}
        I_\text{mono}^\text{inc}(\mathbf{x}) = \sum_\ell [\mathbf{B}]_{\ell \ell}\, W_\ell \;|I_{\ell \ell}(\mathbf{x})|     
\end{equation}
\begin{figure}[!t]
\centering
\subfloat[][Incoherent combination]{\includegraphics[width=0.5\columnwidth]{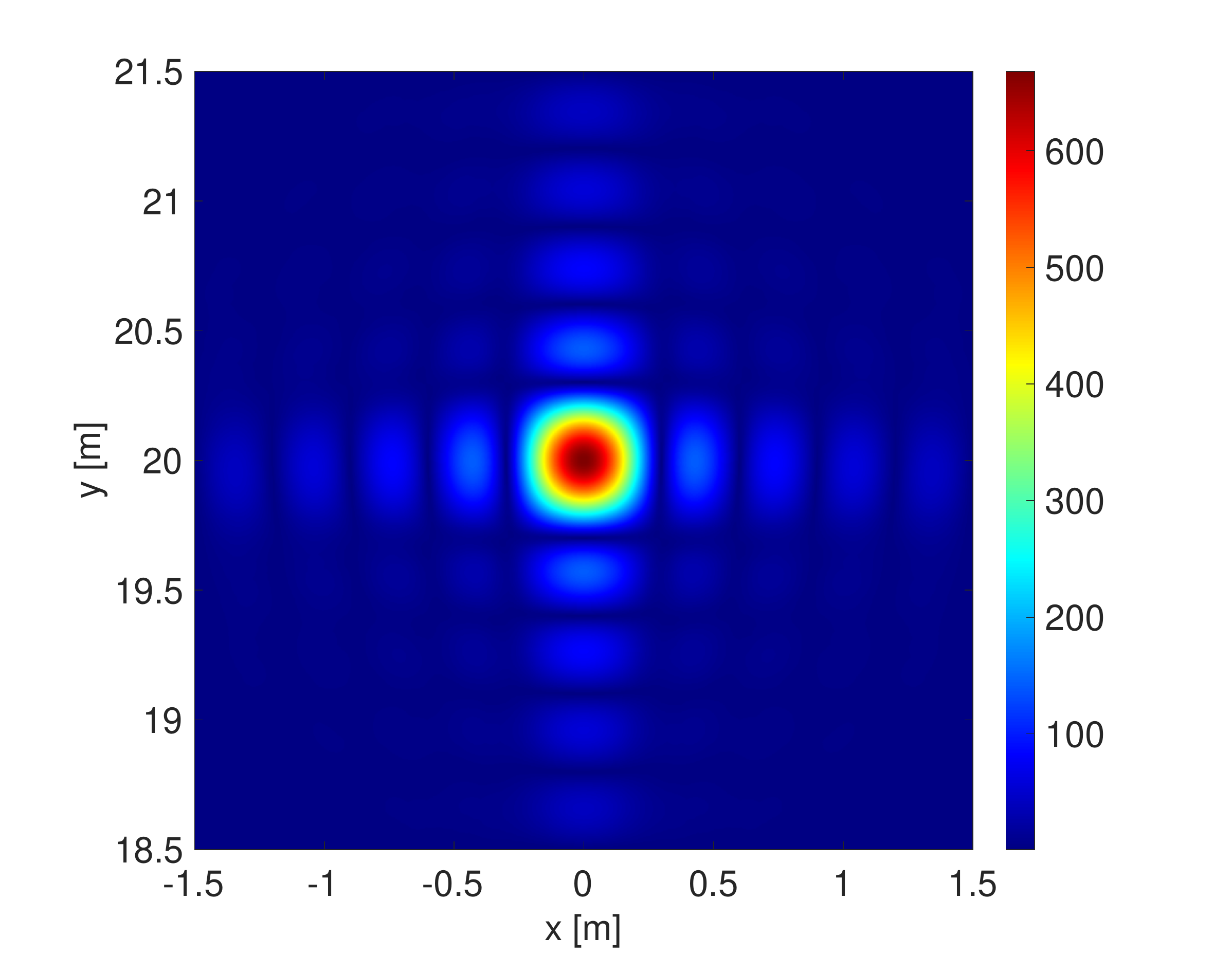}\label{subfig:incoherent_average}} 
\subfloat[][Coherent combination]{\includegraphics[width=0.5\columnwidth]{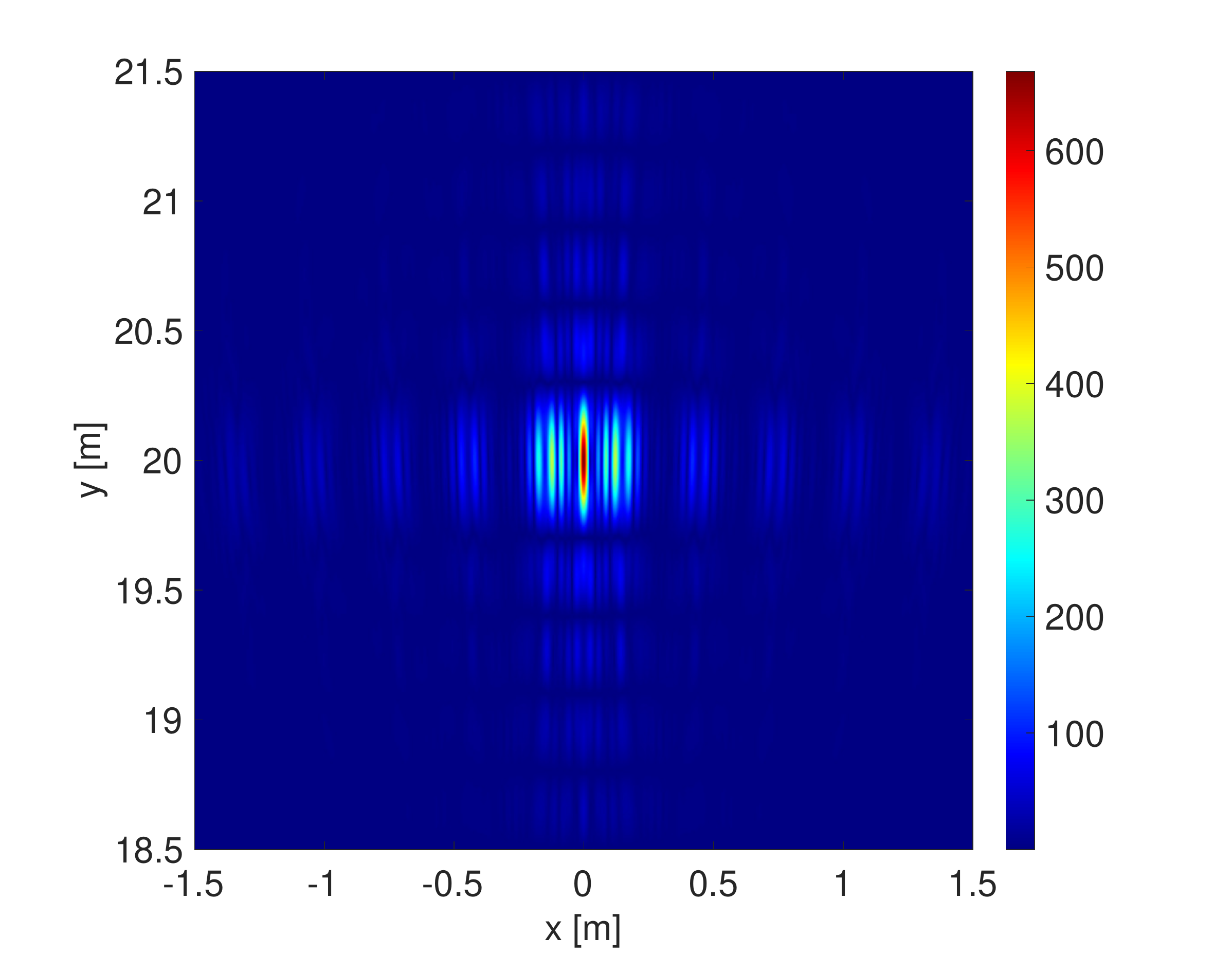}\label{subfig:coherent_average}}\\
\subfloat[][Incoherent spectral coverage]{\includegraphics[width=0.5\columnwidth]{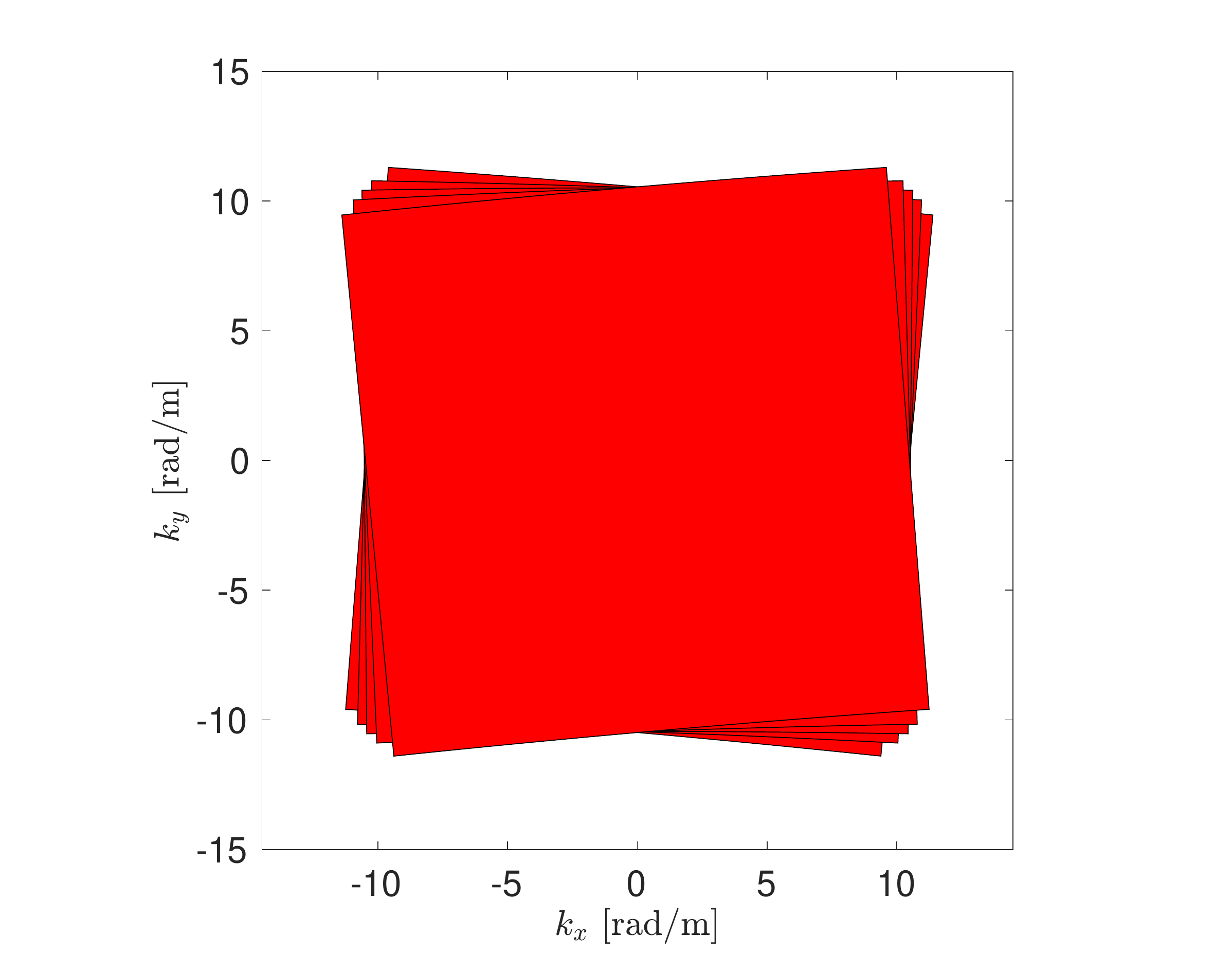}\label{subfig:wavenumber_domain_monostatic_incoherent}} 
\subfloat[][Coherent spectral coverage]{\includegraphics[width=0.5\columnwidth]{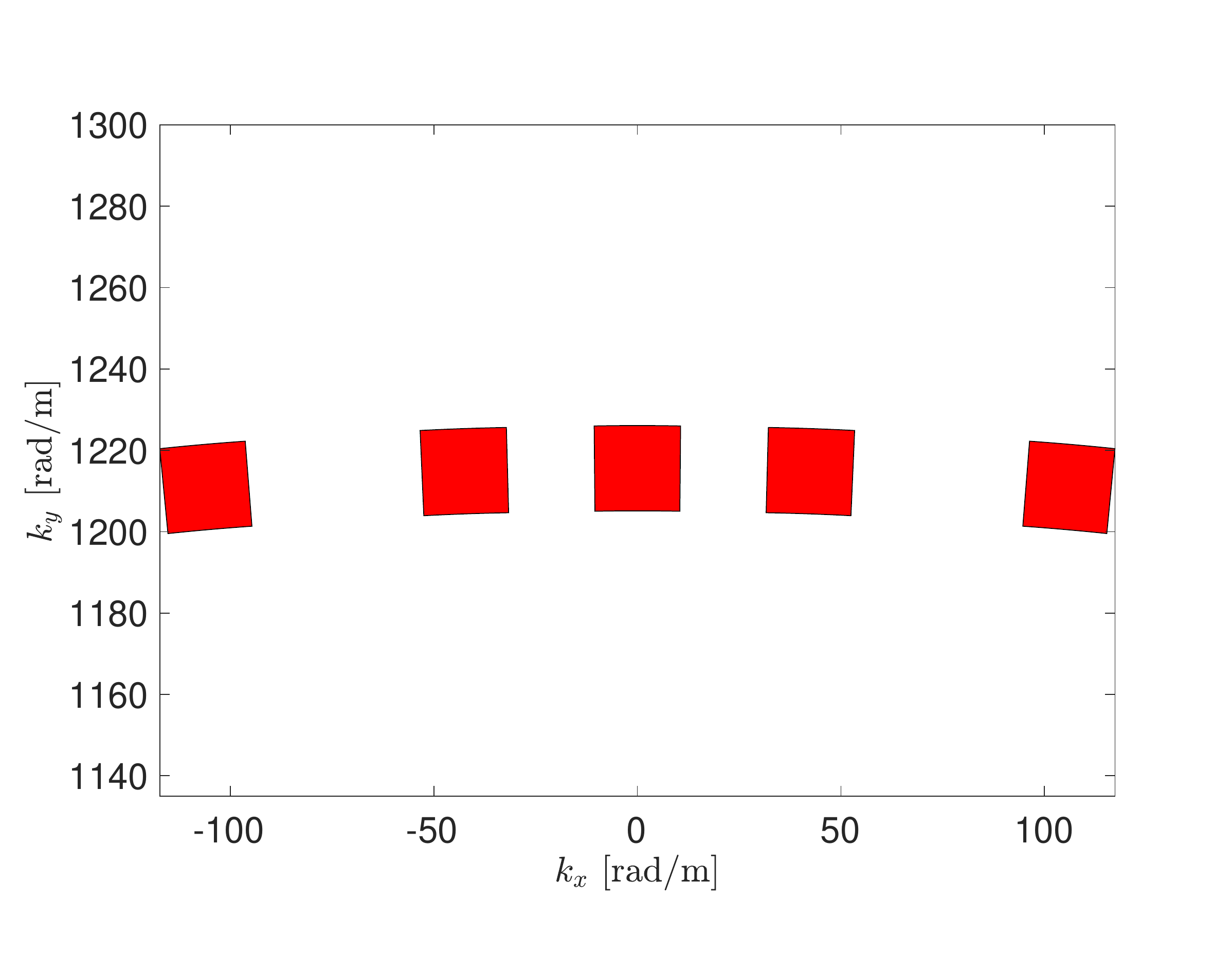}\label{subfig:wavenumber_domain_monostatic_all}}
\caption{Example of imaging of a point target with incoherent and coherent fusion of monostatic images acquired at the 5 terminals in Fig. \ref{fig:geometry_1}. The incoherent fusion is shown in Fig. \ref{subfig:incoherent_average} with corresponding wavenumber covered region in Fig. \ref{subfig:wavenumber_domain_monostatic_incoherent}. The coherent fusion is shown in Fig. \ref{subfig:coherent_average}) with corresponding wavenumber covered region in Fig. \ref{subfig:wavenumber_domain_monostatic_all}.}
\label{fig:mono_coher_incoher}
\end{figure}
where only the absolute value of the image is used and $W_\ell$ is a proper weight for the $\ell$-th image, satisfying some optimality conditions for the combination (e.g., inversely proportional to the quality of the single image, in terms of resolution, SNR, etc.). The only requirement is that the cooperating terminals know each other position (either absolute or relative, as detailed at the end of Section \ref{sec:TDBP_section}) with an accuracy in the order of the spatial resolution, thus centimeter or tens of centimeters for typical systems. No mutual clock/phase synchronization is enforced in this case. Notice that taking the absolute value of each image means mapping the pass-band wavenumber covered region $\mathcal{K}_{B}$ from \eqref{eq:wavenumber_covered_region_aperture} in Section \ref{subsect:monostatic_wavenumber_region}, into the base-band one $\widetilde{\mathcal{K}}_{B}$ which is translated to the origin of the $k_x k_y$ plane, as defined by \eqref{eq:monostatic_comb_coverage_incoherent}. In this case, as illustrated in Figs. \ref{subfig:incoherent_average} and \ref{subfig:wavenumber_domain_monostatic_incoherent}, the overall image $I_\text{mono}^\text{inc}(\mathbf{x})$ of the targets does not improve in resolution, because the overall wavenumber covered region is the same. Under the assumption that the amplitude image formed by each entity is the sum of the true amplitude of the target and additive noise, the gain of such a cooperation is in the SNR, that linearly improves with the number of images incoherently summed (thus by $L$ times if all the terminals cooperate).  
\subsubsection{Coherent sum of images} The coherent combination of monostatic images is a further step in terms of performance and complexity: 
\begin{equation}\label{eq:coherent_sum_mono}
        I_\text{mono}^\text{coh}(\mathbf{x}) = \sum_\ell [\mathbf{B}]_{\ell \ell} \, W_\ell \; I_{\ell \ell}(\mathbf{x}).    
\end{equation}
where, now, the cooperating terminals shall be clock/phase synchronized to properly fuse the complex-valued (amplitude+phase) images and avoid image blurring (\textit{defocusing}). Moreover, the positioning requirements are tighter, i.e., terminals shall know the location (and the orientation in space) of all the others within an accuracy compared with the carrier wavelength $\lambda_0=c/f_0$. A reasonable degree of clock synchronization might be achieved by using GPS-disciplined oscillators (GPSDO) \cite{Beasley2022}. Another solution is to use data-driven methods exploiting the redundancy of the acquired data to estimate and compensate for timing errors between nodes \cite{Azcueta2017}.
Differently from \eqref{eq:incoherent_sum}, in \eqref{eq:coherent_sum_mono} the phase information is preserved and can be used to not only improve the SNR but also the resolution of the generated image $I_\text{mono}^\text{coh}(\mathbf{x})$. The effect of a coherent combination of sensing data can be appreciated in Figs. \ref{subfig:coherent_average} and \ref{subfig:wavenumber_domain_monostatic_all}. In this case, the overall covered region in the wavenumber domain is substantially enlarged compared to the previous case, thus the main lobe of the generated image (around the target's position) is strongly narrowed. The image has gained resolution, but, still, non-negligible grating lobes appear around the main one, in accordance with the sampling theorem applied to the wavenumber domain. For fixed sensing network geometry and monostatic acquisitions, however, this result is the upper bound. Tangible improvements can be attained by filling the unwanted spectral ''holes'' in Fig. \ref{subfig:wavenumber_domain_monostatic_all} by means of multistatic acquisitions, as described in the following.


\subsection{Fusion of Multistatic Images}
\label{sec:multistatic}

\begin{figure}[!t]
\centering
\subfloat[][]{\includegraphics[width=0.55\columnwidth]{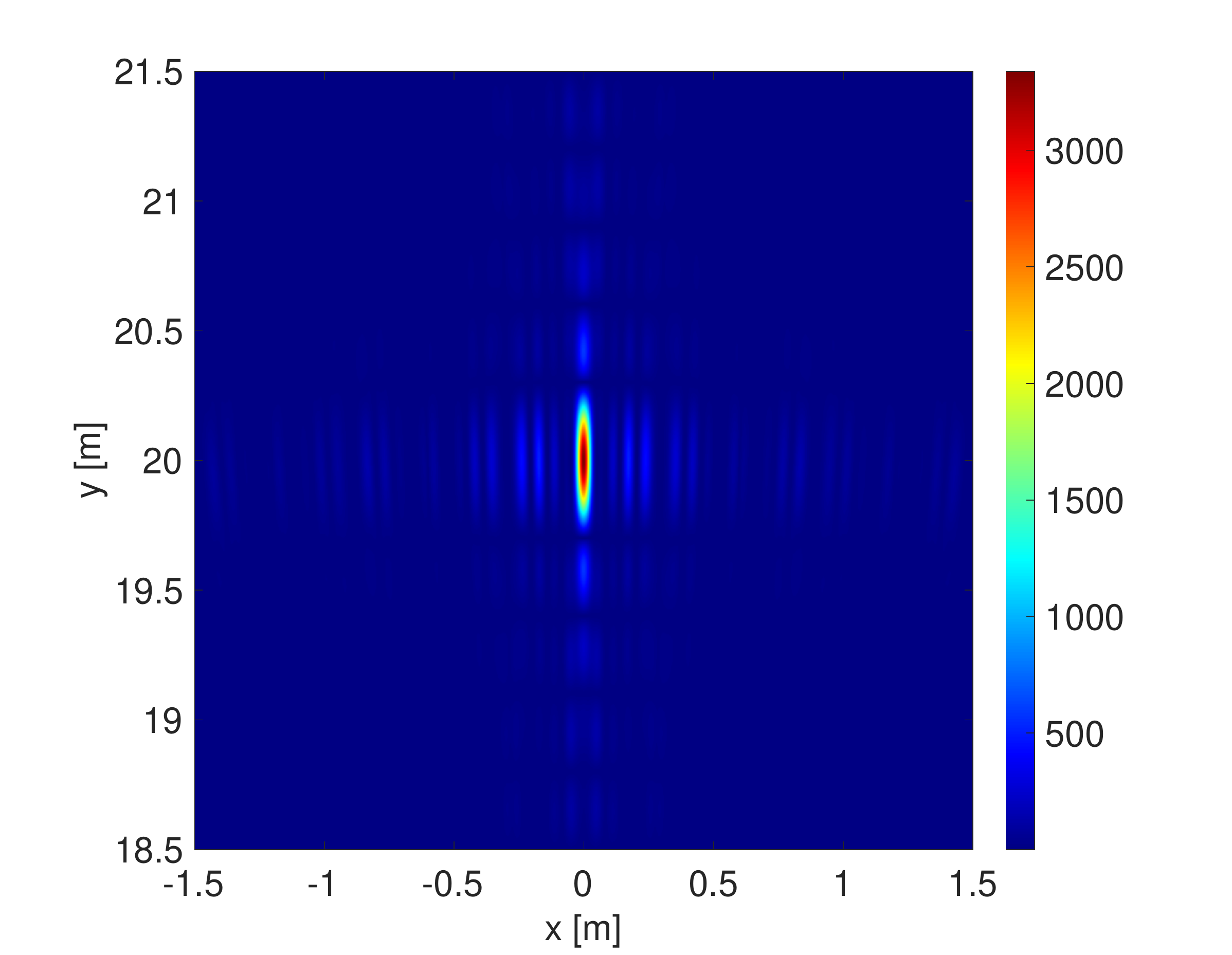}\label{subfig:coherent_average_multistatic}}
\subfloat[][]{\includegraphics[width=0.5\columnwidth]{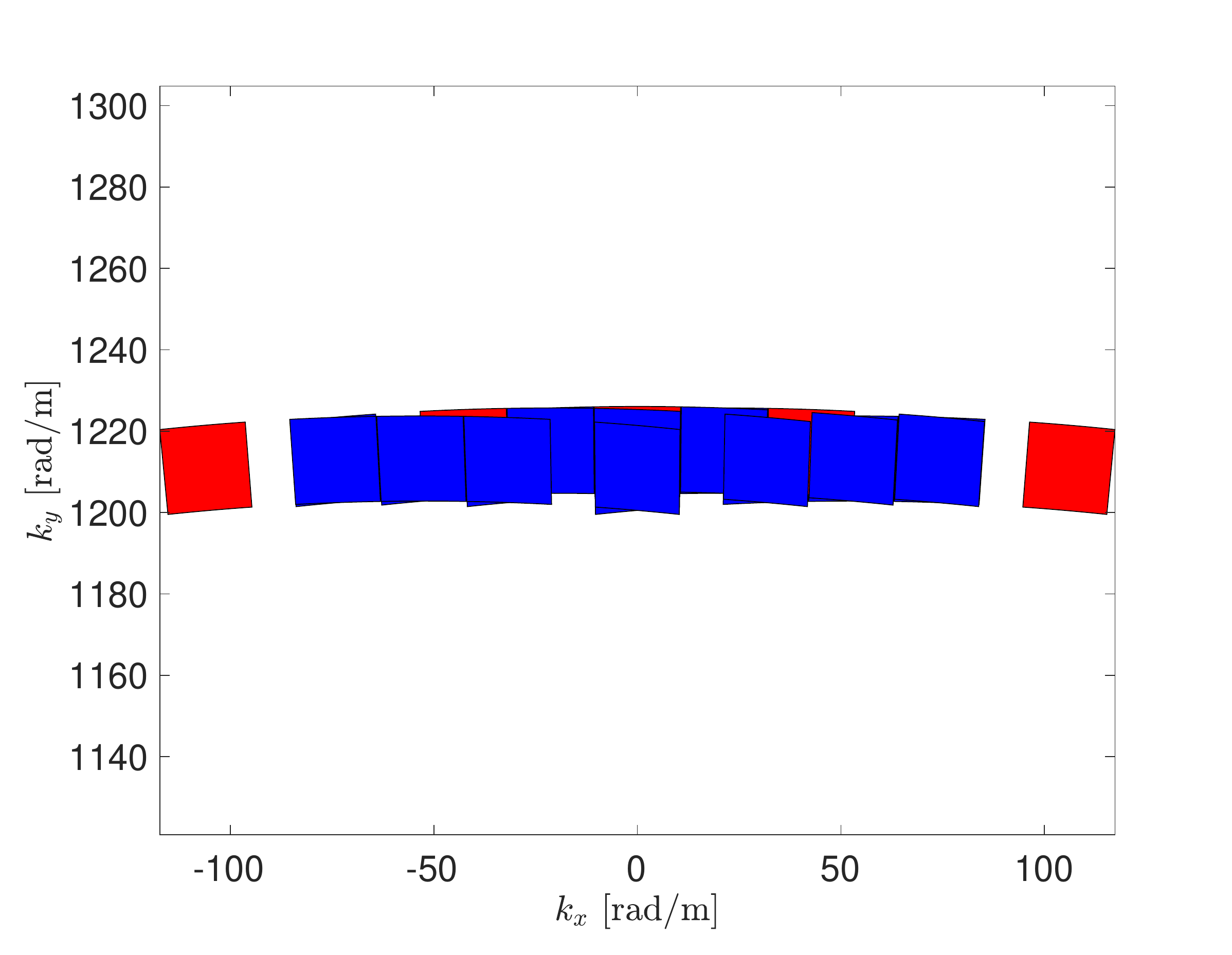}\label{subfig:wavenumber_domain_multistatic_all}} 
\caption{Example of imaging of a point target with coherent fusion of both monostatic and multistatic images acquired at the single terminals. Fig. \ref{subfig:coherent_average_multistatic} shows the image, and the related wavenumber coverage is shown in Fig. \ref{subfig:wavenumber_domain_multistatic_all}. Red tiles represent the spectral support of monostatic acquisitions, the blue tiles represent the additonal coverage brought by multi-static acquisitions.}
\label{fig:coher_multistatic}
\end{figure}

A further step in cooperation is to consider multistatic acquisitions and coherent fusion. In the most general case, all the terminals receive the scattering echoes from the environment pertaining to Tx signal by all the terminals. Thus, we have up to $L^2$ images to coherently combine after being exchanged over a communication network. The result is:
\begin{equation}
\label{eq:coherent_sum}
        I^\text{coh}_\text{multi}(\mathbf{x}) = \sum_{\ell} \sum_k [\mathbf{B}]_{\ell k} \, W_{\ell k}\; I_{\ell k}(\mathbf{x}).
\end{equation}
This configuration is by far the best one in terms of image quality for both resolution and grating lobes suppression. Notice that, in this case, \textit{all} the multistatic Tx-Rx pairs shall be phase-synchronized to enable the fusion as for \eqref{eq:coherent_sum}. An example of multistatic imaging is the one in Fig. \ref{subfig:coherent_average_multistatic}. In this scenario, all the $L$ terminals transmit and receive. Once obtained the single images with back-projection in \eqref{eq:BP}, and fused with \eqref{eq:coherent_sum}, the image of the target shows very fine resolution, high peak value due to the coherent combination of $L^2$ images (instead of $L$ as in the previous example) and reduced grating lobes. The latter is a direct consequence of the wavenumber coverage (Fig. \ref{subfig:wavenumber_domain_multistatic_all}): the gaps in the spectrum in Fig. \ref{subfig:wavenumber_domain_monostatic_all} are now filled by new multistatic spectrum tiles.
\subsection{Orchestration: the Wavenumber Tessellation Principle}
\label{sec:orchestration}
From the example in the previous Section \ref{sec:multistatic}, it is clear that utter imaging performance can be attained by exploiting multistatic acquisitions. However, this implies synchronizing the intended terminals and estimating their relative position in space with high accuracy, e.g., by using multiple positioning sensors \cite{Manzoni2022_TITS}. The question is therefore: is it possible to achieve the approximately same imaging quality of Fig. \ref{fig:coher_multistatic} by enforcing the cooperation of a reduced set of terminals ($< L$)? The answer lies in the concept of \textit{orchestrated sensing}. In the previous example (Section \ref{sec:multistatic}), all the sensors are both Tx and Rx, to create the $L^2$ combinations maximizing the performance. Orchestrated sensing evolves from this concept allowing one to choose the set of cooperating terminals and the allocated resources (e.g., employed bandwidth, power, time, and space) to fulfill a pre-defined goal. Therefore, orchestration differs from simple cooperation in that sensing resources are allocated to each intended terminal to maximize some target figure of merit for sensing. In our case, we use the DTT to maximize the resolution of the image as well the grating lobes suppression pursuing the \textit{wavenumber tessellation} principle.

Let us consider an example of orchestration where the $L$ terminal can move along the $x$ axis, e.g., vehicles traveling along the lane. Now, we assume that every single terminal employs a bandwidth $B=100$ MHz (e.g. for ISAC systems operating in the 5G NR FR1 spectrum portion), thus the range resolution of the previous examples cannot be achieved for the geometry in Fig. \ref{fig:geometry_1}, even in the case of a full multistatic acquisition, as the wavenumber coverage is large along $k_x$ and not along $k_y$. The wavenumber tessellation principle allows choosing \textit{when} and \textit{where} each terminal operates such that to increase the wavenumber coverage along  $k_y$, still without grating lobes. It can be proved that, if the observation angles $\{\psi_\ell\}_{\ell=1}^L$ are chosen according to \cite{Tebaldini2017_tessellation}:
\begin{equation}
    \sin(\psi_\ell) = \sin(\psi_{\ell-1})\left(f_0-\frac{B}{2}\right)\left(f_0+\frac{B}{2}\right)^{-1}  
\end{equation}
the wavenumber tiles for each multistatic acquisition are contiguous and thus the resolution along $y$ is $L$-times larger compared to the single terminal case ($L$-fold increase in the equivalent employed bandwidth).

\begin{figure}[!t]
\centering
\subfloat[][Single terminal]{\includegraphics[width=0.5\columnwidth]{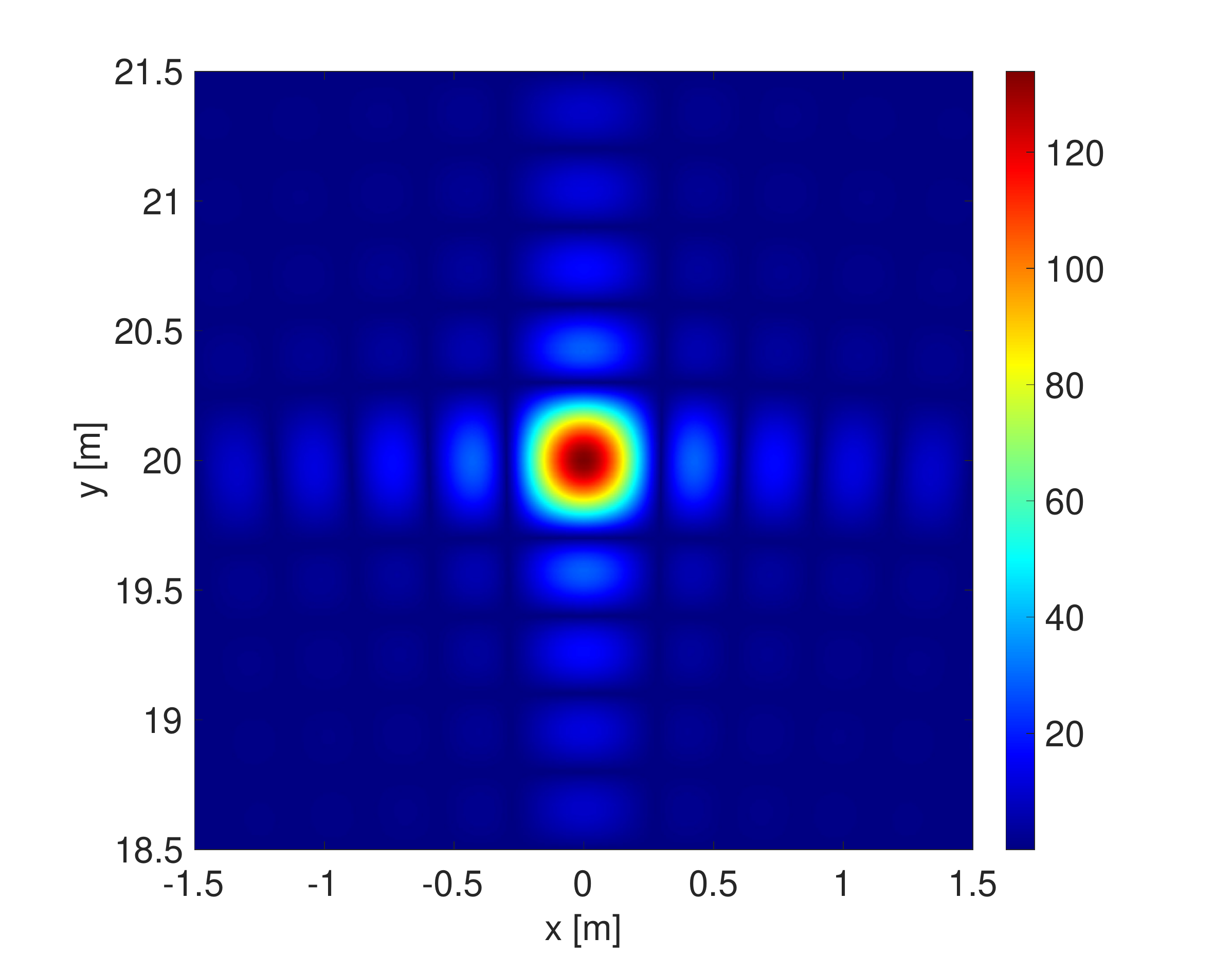}\label{subfig:single_node_orchestration}} 
\subfloat[][Orchestration]{\includegraphics[width=0.5\columnwidth]{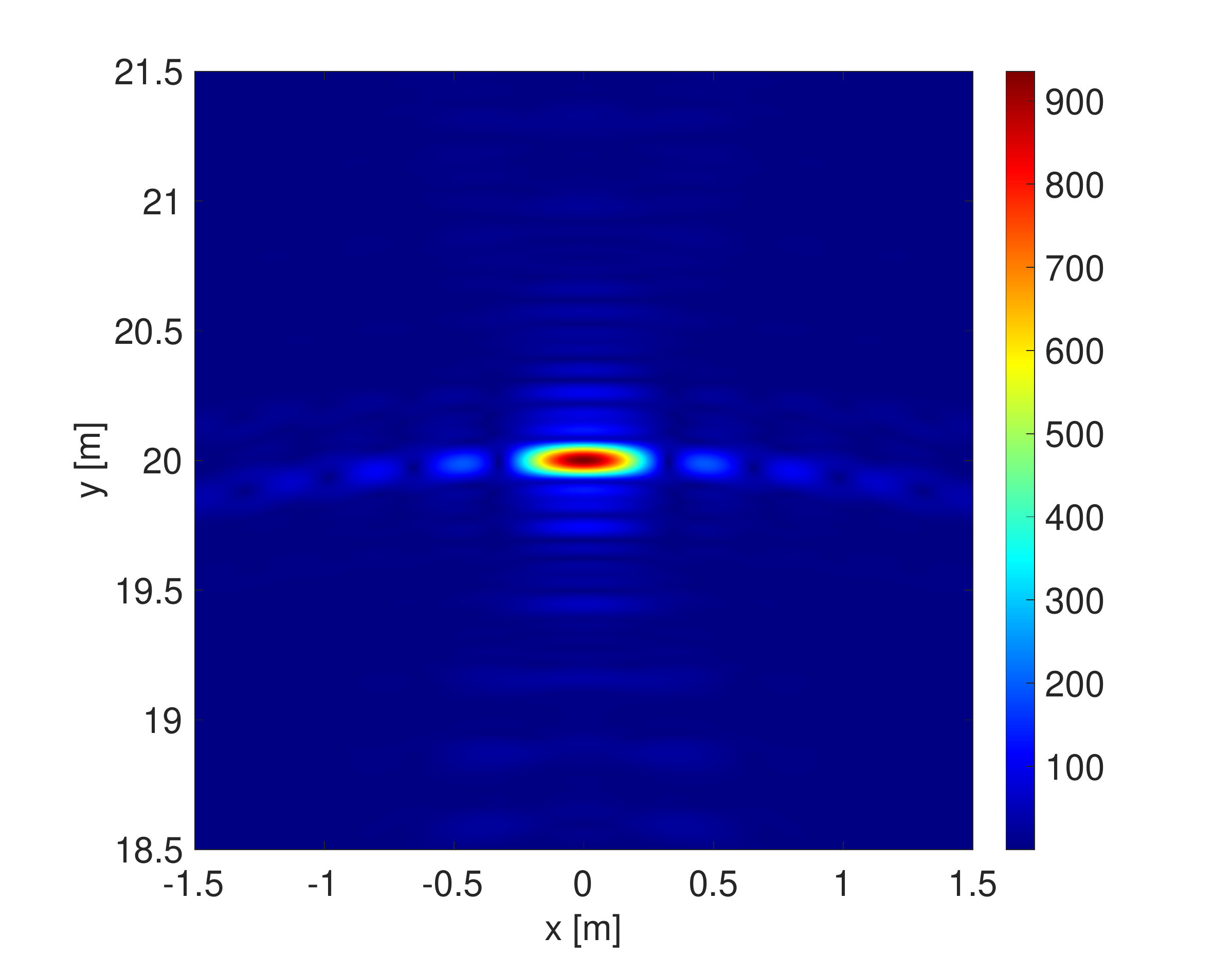}\label{subfig:orchestration}}\\
\subfloat[][Single terminal spectral coverage]{\includegraphics[width=0.5\columnwidth]{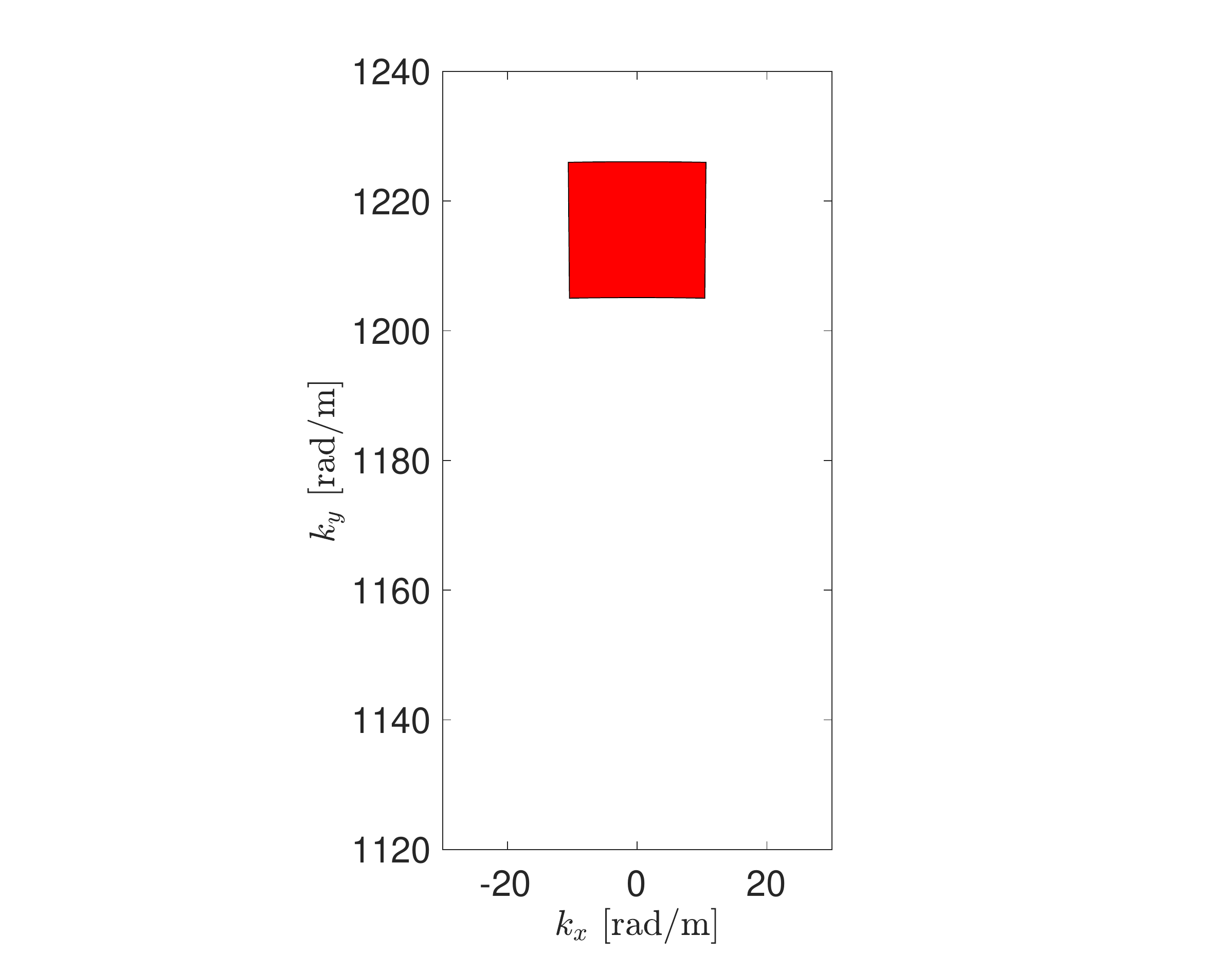}\label{subfig:wavenumber_domain_orchestration_single}} 
\subfloat[][Orchestration spectral coverage]{\includegraphics[width=0.5\columnwidth]{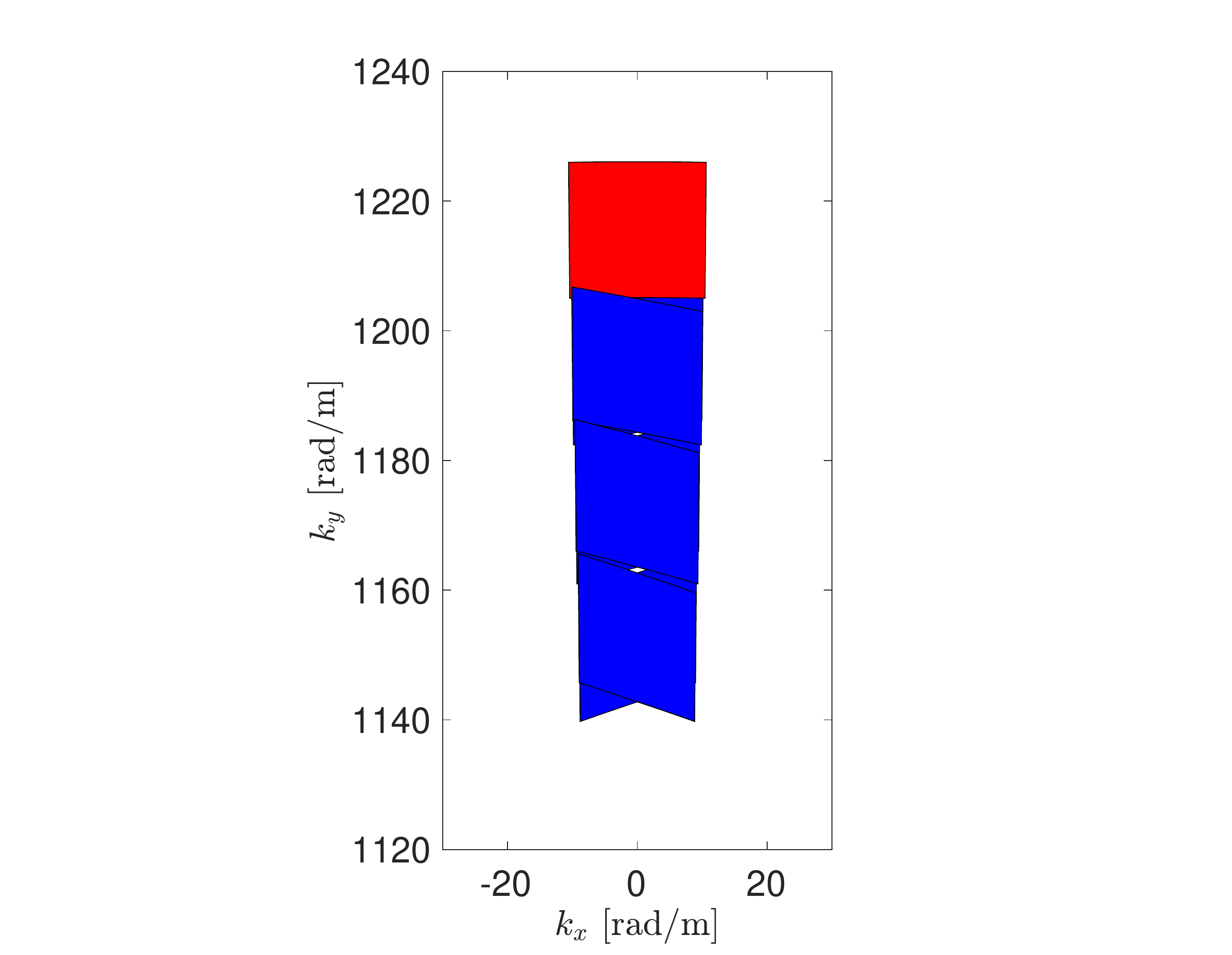}\label{subfig:wavenumber_domain_orchestration}}

\caption{Example of orchestrated sensing for the imaging of a point target. Fig. \ref{subfig:single_node_orchestration} shows the single-terminal image (with related wavenumber coverage in Fig. \ref{subfig:wavenumber_domain_orchestration_single}), while the result of the wavenumber tessellation is in Figs. \ref{subfig:orchestration} and \ref{subfig:wavenumber_domain_orchestration}. Red tiles represent the wavenumber coverage of monostatic acquisitions, while blue tiles the multi-static counterpart. The resulting image has four times the resolution along $y$, meaning a $4\times$-fold increase in the equivalent employed bandwidth.}
\label{fig:orchiestration_main_figure}
\end{figure}

The image generated by a single terminal of such a system is depicted in Fig. \ref{subfig:single_node_orchestration} and its wavenumber domain coverage is portrayed in Fig. \ref{subfig:wavenumber_domain_orchestration_single}. Once again, the aperture of each terminal is set in such a way as to have equal resolutions along the $x$ and $y$ directions. The result of the orchestration is portrayed in Fig. \ref{subfig:wavenumber_domain_orchestration}. The resolution along the $y$ coordinate is four times better than the previous one, without significant grating lobes, which is a direct consequence of the spectrum broadening with contiguous tiles in Fig. \ref{subfig:wavenumber_domain_orchestration}. The result is remarkable: even in limited bandwidth conditions, the resolution can be enhanced following a simple principle driven by the DTT, that does only require the knowledge of the acquisition geometry and sensing capabilities of each terminal. 

\section{An Example of Design Guideline }\label{sec:design_guideline}
\begin{figure}[!t]
    \centering
    \includegraphics[width=0.8\columnwidth]{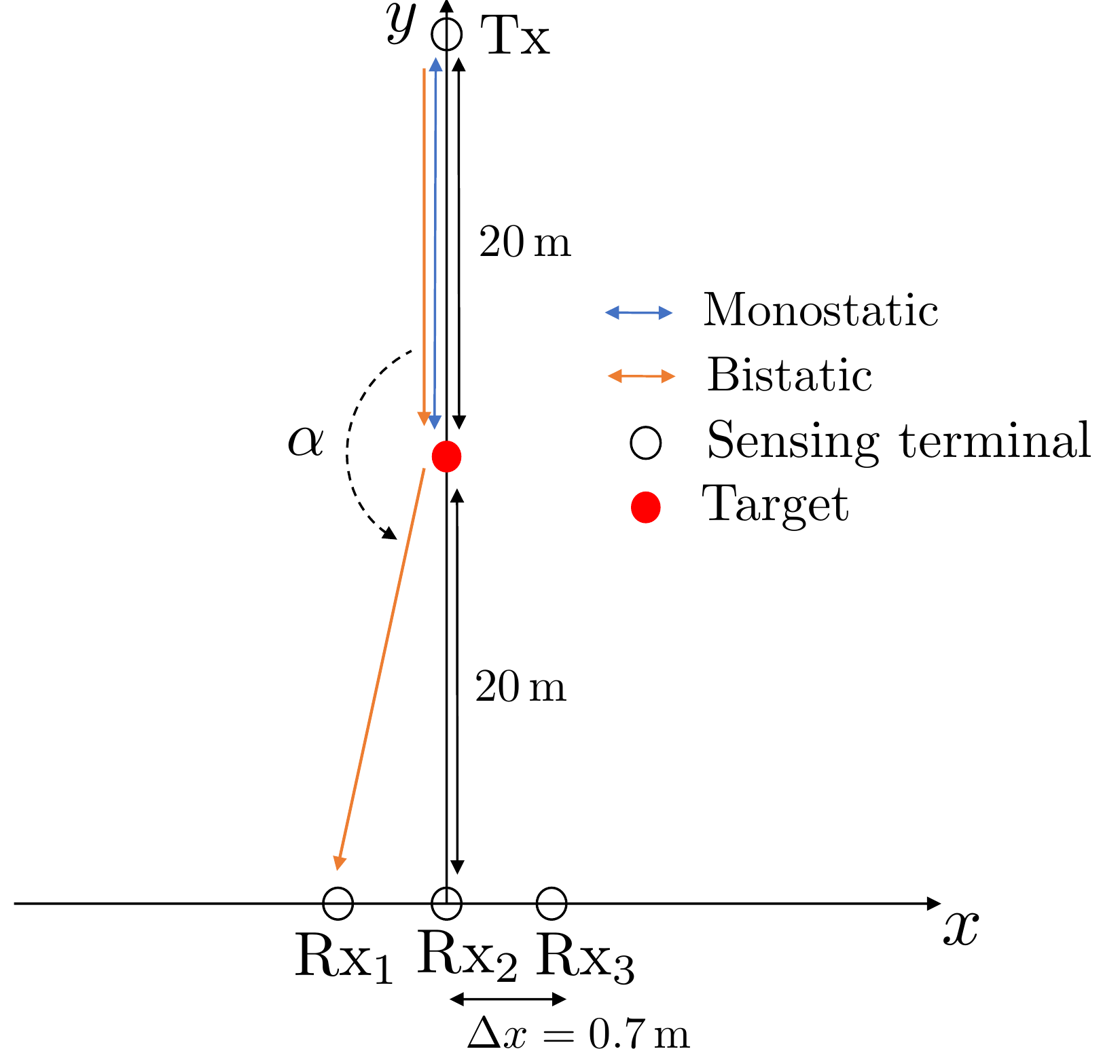}
    \caption{Considered sensing network topology (image not in scale).}
    \label{fig:geometry_2}
\end{figure}

The previous section outlined the overall resolution performance that can be attained when considering multistatic acquisitions and coherent combination of images. We herein provide an application example for networked sensing systems where the cooperation among terminals is not recommended. We choose as a use case the scenario in Fig. \ref{fig:geometry_2}, showing one Tx terminal, e.g., an ISAC BS, is located in in $\mathbf{x}_\mathrm{Tx} = [0,40]^\mathrm{T}$ m and multiple Rx terminals, e.g., users, are displaced along the $x$ axis spaced by $\Delta x = 0.7$ m. The intended target to be imaged is between Tx and Rx, in $\mathbf{x}_t = [0,20]^\mathrm{T}$ m.This could be the case of multiple vehicles traveling on a lane (the Rxs) imaging the environment underneath the BS/RSU (the Tx), using the emitted communication signal by the latter. The employed bandwidth is rather large, $B=500$ MHz.
\begin{figure*}[!t]
\centering
\subfloat[][Monostatic]{\includegraphics[width=0.55\columnwidth]{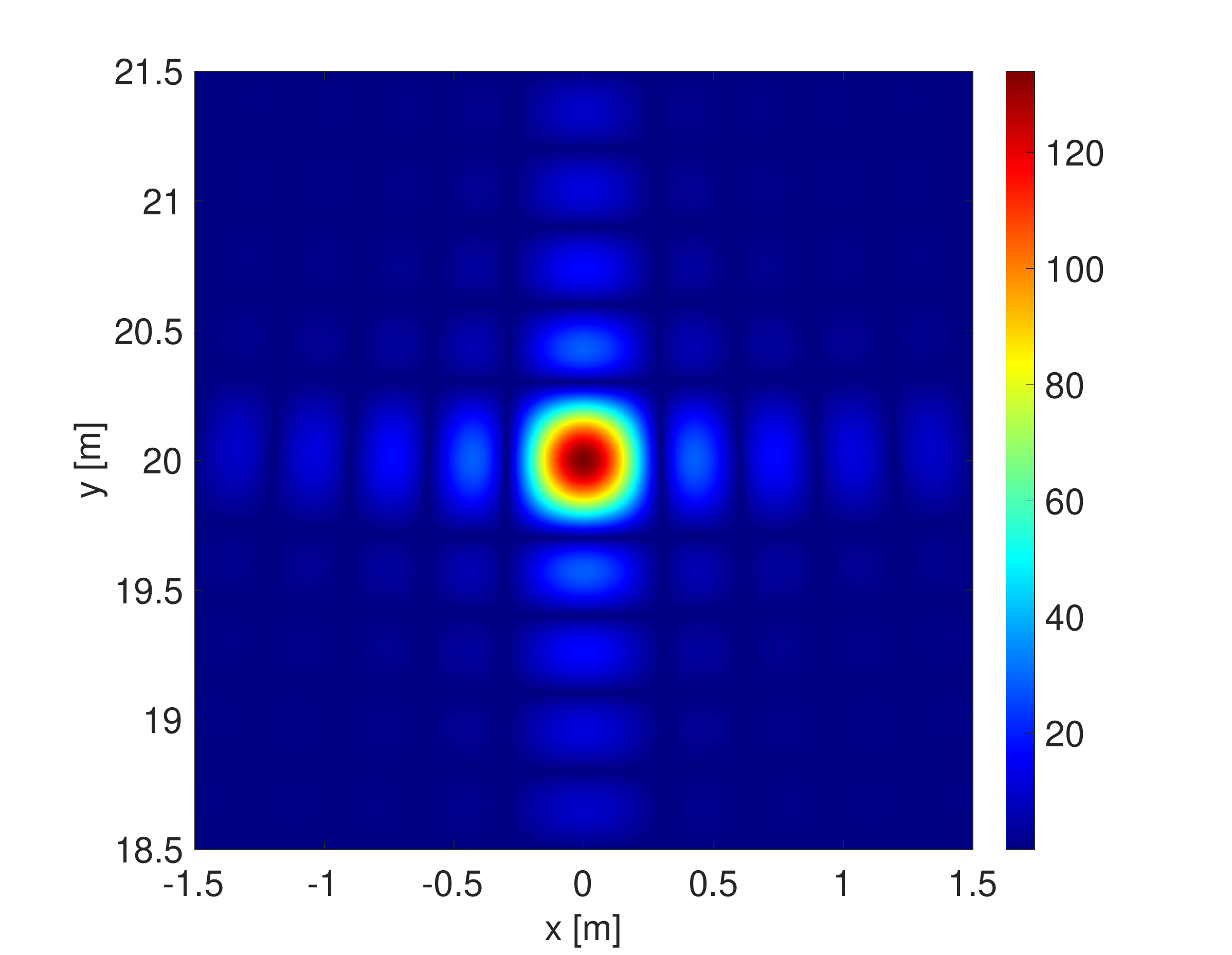}\label{subfig:monostatic_opposite}}
\subfloat[][Bistatic]{\includegraphics[width=0.55\columnwidth]{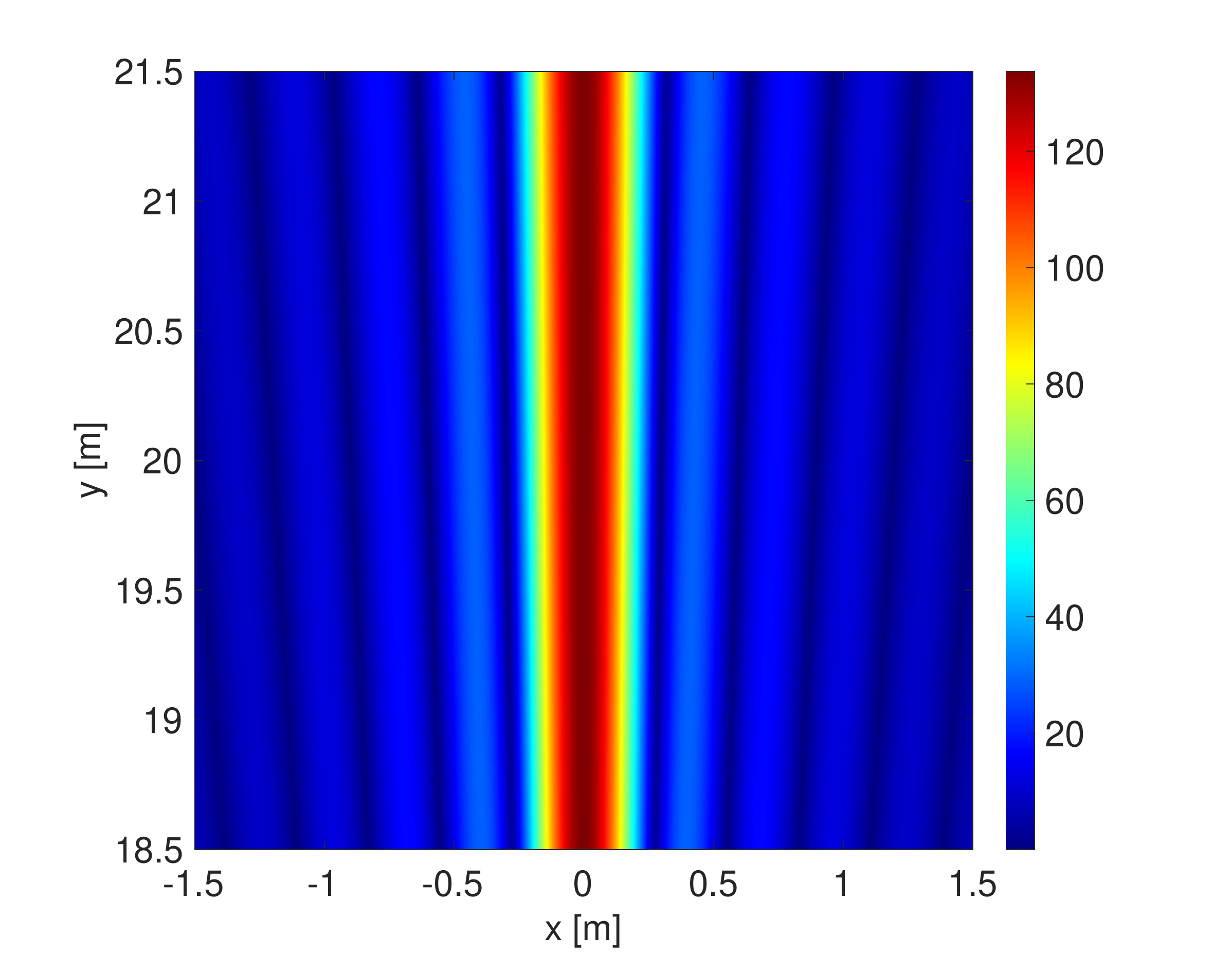}\label{subfig:bistatic_opposite}}
\subfloat[][Multistatic coherent fusion]{\includegraphics[width=0.55\columnwidth]{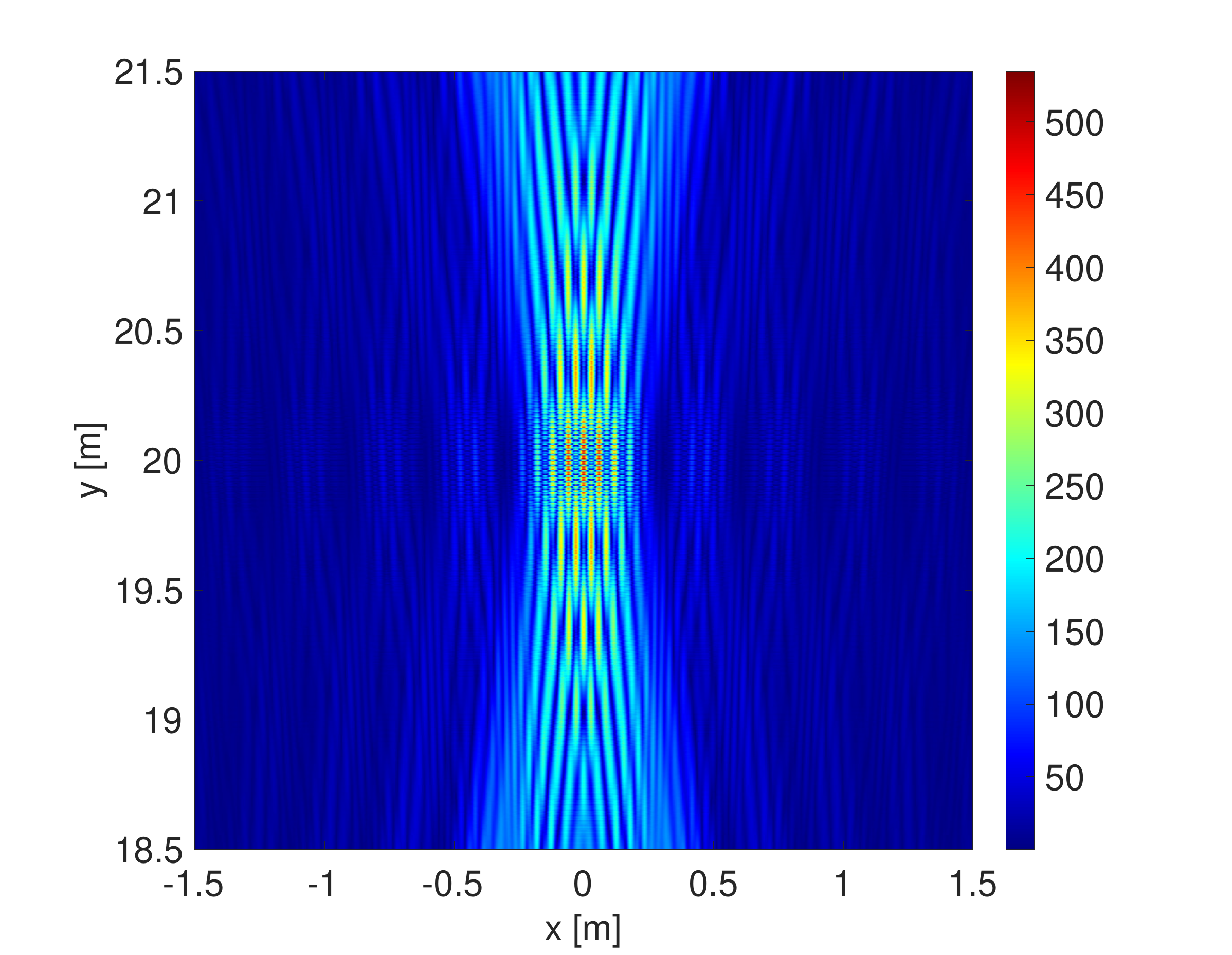}\label{subfig:bistatic_opposite_all}}\\
\subfloat[][Monostatic spectral coverage]{\includegraphics[width=0.55\columnwidth]{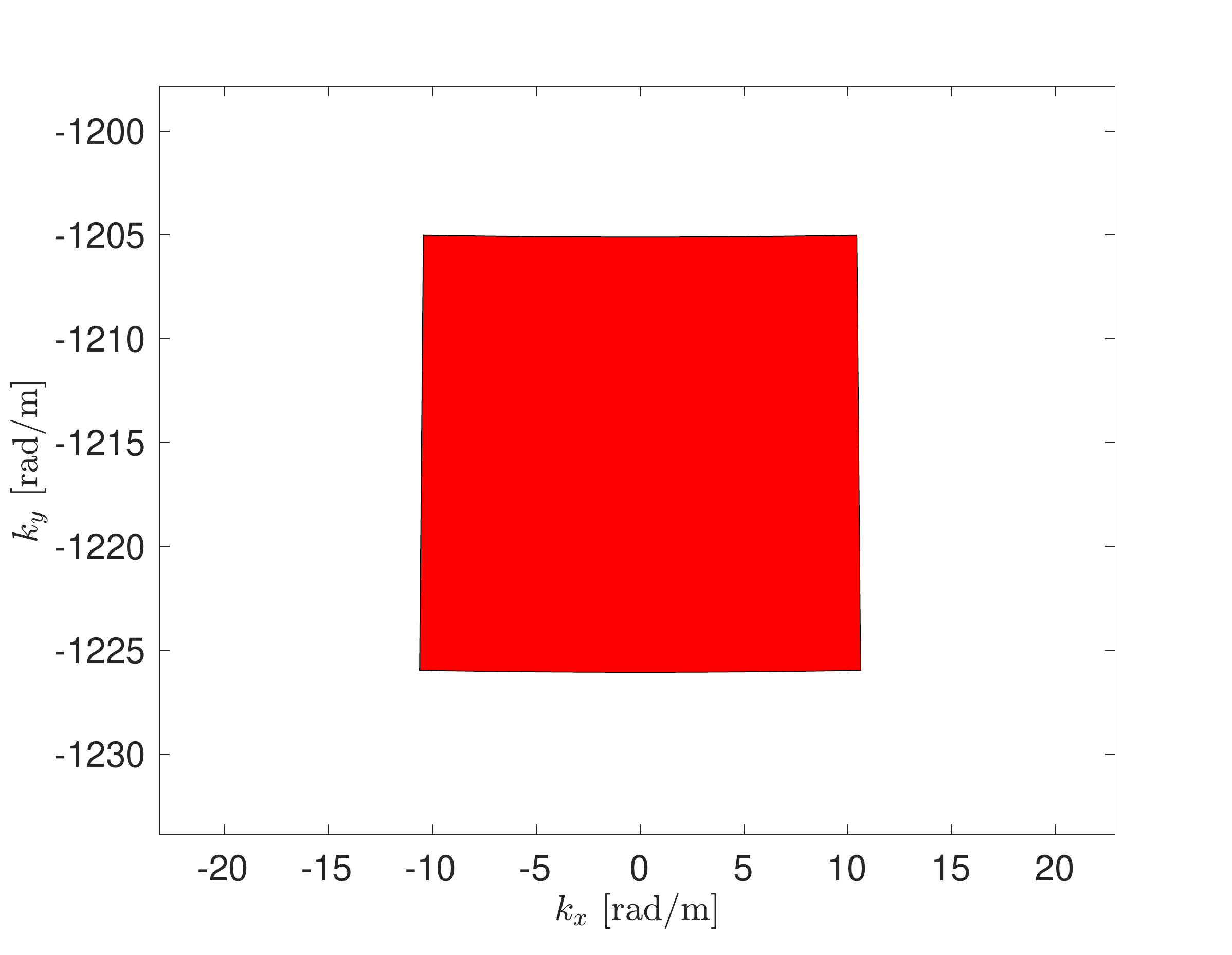}\label{subfig:wavenumber_domain_opposite_side_mono}}
\subfloat[][Bistatic spectral coverage]{\includegraphics[width=0.55\columnwidth]{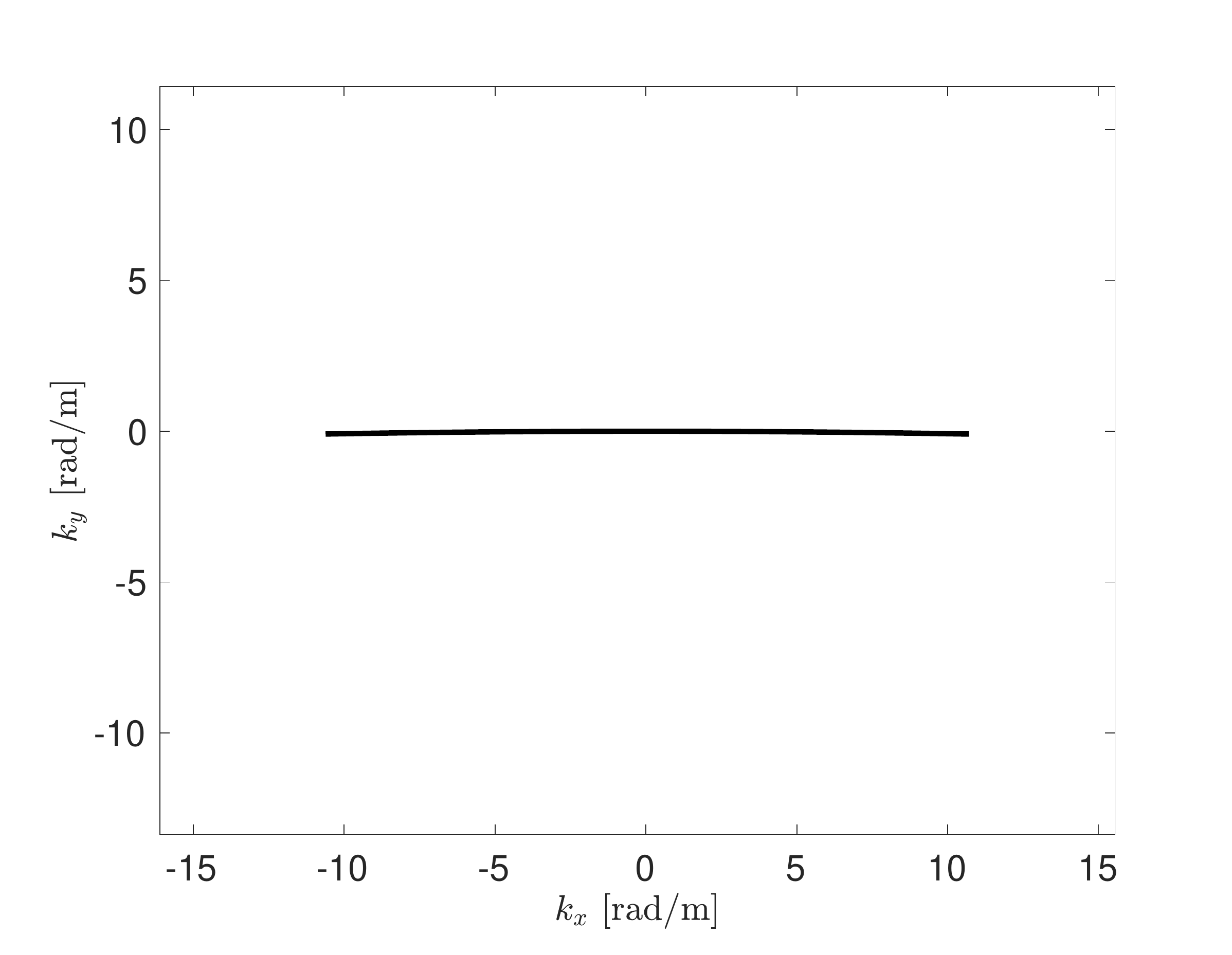}\label{subfig:wavenumber_domain_opposite_side}}
\subfloat[][Multistatic coherent spectral coverage]{\includegraphics[width=0.55\columnwidth]{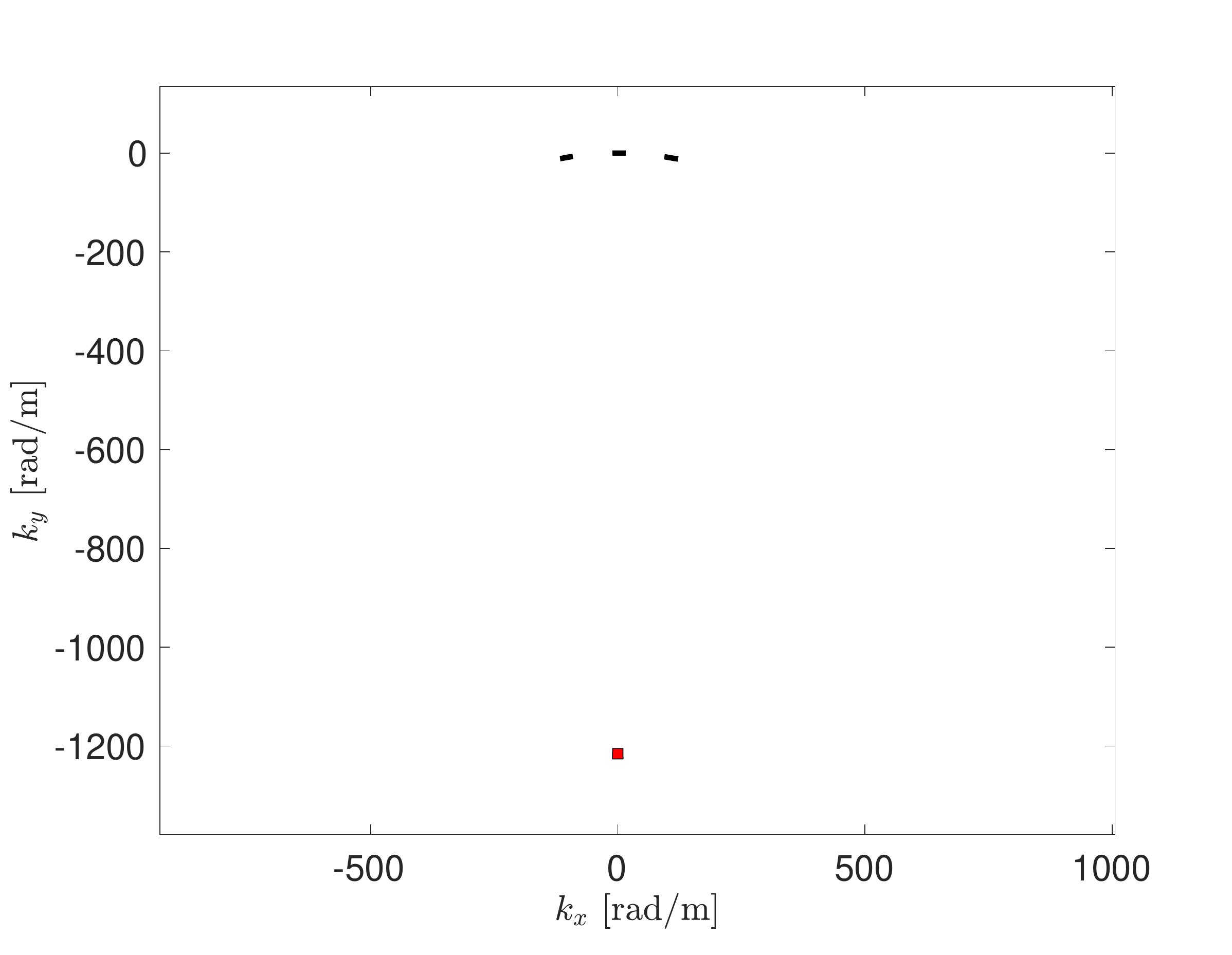}\label{subfig:wavenumber_domain_opposite_side_all}}
\caption{Example of imaging of a point target in between Tx and Rx terminals. Figs. \ref{subfig:monostatic_opposite} and \ref{subfig:wavenumber_domain_opposite_side_mono} show the monostatic image at the Tx terminal and its spectral content. Figs. \ref{subfig:bistatic_opposite} and \ref{subfig:wavenumber_domain_opposite_side} show the bistatic image at the Rx2 side and related wavenumber coverage, while Figs. \ref{subfig:bistatic_opposite_all} and \ref{subfig:wavenumber_domain_opposite_side_all} show the effct of the coherent combination. The specific geometry does not allow to have satisfactory resolution along the $y$ axis, as there is no spectral coverage. Moreover, the wavenumber spectrum is too sparse to enable a fruitful cooperation between tarminals. }
\label{fig:opposite_side}
\end{figure*}
Fig. \ref{subfig:monostatic_opposite} shows the monostatic image obtained by the Tx/Rx terminal. This is limited in resolution by the available bandwidth and number of antennas of the single terminal. Fig. \ref{subfig:bistatic_opposite}, instead, shows the image associated with the Tx-Rx3 bistatic couple of Fig. \ref{fig:geometry_2}. The image has no resolution along $y$. The motivation for this unsatisfactory behavior of bistatic pairs can be easily inferred from the wavenumber-covered region in Fig. \ref{subfig:wavenumber_domain_opposite_side}. According to the DTT, the spectral coverage of a sensing acquisition between the $\ell$-th and $k$-th terminal is formed by the collection of composite wavevectors $\mathbf{k}^* = \mathbf{k}_{\mathrm{Tx},n}^\ell - \mathbf{k}_{\mathrm{Rx },m}^k$ (for each frequency $f\in[f_0-(B/2),f_0+(B/2)]$ and measurement channel $n m$, see \eqref{eq:wavenumber_region_segment} in Section \ref{sect:FEDT}). In the considered case, i.e., the Tx-Rx3, the Tx and Rx wavevectors have \textit{opposite} directions along $k_y$, giving rise to \textit{no coverage} (thus no resolution) in $y$. In other words, having a large bandwidth tends to be useless. A similar consideration applies, with due modifications, to all the other Tx-Rx pairs. It can be easily demonstrated with simple geometrical considerations that, given the bistatic angle $\alpha$ defined as in Fig. \ref{fig:geometry_2}, the spatial resolution along $x$ and $y$ is approximated by:
\begin{equation}\label{eq:resolution_loss}
    \rho_x(\alpha) \simeq \frac{\rho_x(0) }{\cos(\alpha/2)},\,\,\,
    \rho_y(\alpha) \simeq \frac{\rho_y(0) }{\cos(\alpha/2)}
\end{equation}
where $\rho_x(0)$ and $\rho_y(0)$ are the monostatic image resolution along $x$ and $y$ defined in Section \ref{subsect:resolution}. The factor $\cos(\alpha/2)$ represents a severe loss of resolution compared to the monostatic case. As a rule of thumb, the loss can be considered as tolerable for $\cos(\alpha/2)>0.5$, i.e., $\alpha \leq 120$ deg. Notice that \eqref{eq:resolution_loss} approximates the resolution by slicing the image along $x$ and $y$, implicitly approximating the bistatic covered region as a rectangle. In practice, defining the resolution is more challenging as the is not separable in global coordinates. These results shed light on the fact that for all bistatic pairs when Tx and Rx are on opposite sides w.r.t. the scene, the \textit{imaging} is not recommended as the performance-cost trade-off is poor and there are no practical benefits. In addition, the resulting spectrum is highly sparse and does not allow to coherently combining the single images to improve the monostatic performance (see Fig. \ref{subfig:bistatic_opposite_all}). The final image is not satisfactory since a large number of grating lobes appear and the position of the target is not clearly detectable. This is directly related to the sparsity o the spatial spectrum depicted in Fig. (\ref{subfig:wavenumber_domain_opposite_side_all}). Remarkably, all the previous considerations and guidelines are obtained from a physics-based theory, that only require basic geometric information about the topology of the sensing network and the target.

\section{Conclusion and Future Perspectives }\label{sect:conclusion}
This paper proposes the usage of DTT as a fundamental physics-driven theory to evaluate networked sensing performance in terms of imaging resolution, where the interest is in the environment mapping. We outline the DTT and derive the back-projection integral from the wavenumber covered region as the most general image formation algorithm. We then discuss the possible options for cooperation, highlighting advantages and disadvantages. The DTT-based wavefield networked sensing theory developed in this paper can be used to interpret and predict the imaging performance of a generic network of sensing terminals, without limitations on the type, sensing capabilities (i.e., resources), and network topology. 
Stemming from this work, several promising research directions can be envisioned. The investigation of the fundamental synchronization requirements among different sensing terminals, and the related effects on the imaging of the target, are of great importance for the identification of practical requirements. Moreover, the orchestration based on wavenumber tessellation can be formally posed as a resource allocation strategy, given the network topology and the sensing capabilities of each terminal. Finally, a deep understanding of the parallelism between DTT and the estimation theory (CRLB) is necessary. On one side, DTT evaluates imaging performance in a noiseless setting, irrespective of the specific Tx signal; on the other side, CRLB considers the specific signal and noise modeling and provides the utter estimation performance for selected target's parameters. The conjunction point between the two theories (if it exists) may lead to developing a novel key performance indicator for sensing.

\bibliographystyle{IEEEtran}
\bibliography{Bibliography}

\end{document}